\documentclass[preprint2]{aastex6}

\newcommand{\Msun}{\ifmmode M_{\odot} \else $M_{\odot}$\fi}
\newcommand{\Msunyr}{\ifmmode M_{\odot}\, {\rm yr}^{-1} \else $M_{\odot}\, {\rm yr}^{-1}$}

\def\sgra{Sgr~A*}
\def\asec{$''$}
\def\spt{$\buildrel{\prime\prime}\over .$}

\def\Chandra{{\it Chandra}}
\providecommand\casa{\textsc{casa}}

\newcommand{\July}{2013 July 27}
\newcommand{\Aug}{2013 August 12}
\newcommand{\Sept}{2013 September 14}
\newcommand{\Oct}{2013 October 28}
\newcommand{\Feb}{2014 February 21}
\newcommand{\Apr}{2014 April 28}
\newcommand{\May}{2014 May 20}
\newcommand{\Julyb}{2014 July 05}

\shorttitle{Sgr A* X-ray and Radio Variability}
\shortauthors{Capellupo et al.}

\begin{document}


\title{Simultaneous Monitoring of X-ray and Radio Variability in Sagittarius A*}


\author{
Daniel M.\ Capellupo$^{1,2}$, Daryl Haggard$^{1,2}$, Nicolas Choux$^{1,2}$,
Fred Baganoff$^{3}$,
Geoffrey C.\ Bower$^{4}$, \\
Bill Cotton$^{5}$,
Nathalie Degenaar$^{6}$,
Jason Dexter$^{7}$,
Heino Falcke$^{8}$,
P.\ Chris Fragile$^{9}$,
Craig O.\ Heinke$^{10}$, \\
Casey J.\ Law$^{11}$,
Sera Markoff$^{6}$,
Joey Neilsen$^{3}$,
Gabriele Ponti$^{7}$,
Nanda Rea$^{6}$,
Farhad Yusef-Zadeh$^{12}$
}
\affil{
$^{1}$Department of Physics, McGill University, Montreal, Quebec, H3A 2T8, Canada \\
$^{2}$McGill Space Institute, McGill University, Montreal, QC, H3A 2A7, Canada \\
$^{3}$MIT Kavli Institute for Astrophysics and Space Research, Cambridge, MA 02139, USA \\
$^{4}$Academia Sinica Institute of Astronomy and Astrophysics, 645 N. A'ohoku Place, Hilo, HI 96720, USA\\
$^{5}$National Radio Astronomy Observatory, Charlottesville, VA 22903, USA \\
$^{6}$Anton Pannekoek Institute for Astronomy, University of Amsterdam, 
Science Park 904, NL-1098 XH Amsterdam, The Netherlands \\
$^{7}$Max Planck Institute for Extraterrestrial Physics, Giessenbachstrasse 1, D-85748 Garching, Germany \\
$^{8}$Department of Astrophysics/IMAPP, Radboud University, PO Box 9010, 
NL-6500 GL Nijmegen, The Netherlands \\
$^{9}$Physics and Astronomy College of Charleston, 66 George Street, 
Charleston, SC 29424, USA \\
$^{10}$Department of Physics, CCIS 4-183, University of Alberta, Edmonton, AB 
T6G 2E1, Canada \\
$^{11}$Radio Astronomy Lab and Department of Astronomy, 501 Campbell Hall, University of California, Berkeley, CA 94720, USA \\
$^{12}$Department of Physics and Astronomy and CIERA, Northwestern University, Evanston, IL 60208, USA
}




\altaffiltext{1}{danielc@physics.mcgill.ca}

\begin{abstract}
Monitoring of Sagittarius~A* from X-ray to radio wavelengths has revealed
structured variability --- including X-ray flares --- but it is challenging
to establish correlations between them. Most studies have
focused on variability in the X-ray and infrared, where variations are often
simultaneous, and because long time series at sub-millimeter and radio
wavelengths are limited.
Previous work on sub-mm and radio variability hints at a lag between 
X-ray flares and their candidate sub-millimeter or radio counterparts, with
the long wavelength data lagging the X-ray.
However, there is only one published time lag between an X-ray flare and a 
possible radio counterpart.
Here we report 9 contemporaneous X-ray and radio observations of \sgra.
We detect significant radio variability peaking $\gtrsim$176 minutes after
the brightest X-ray flare ever detected from \sgra. We also report other
potentially associated X-ray and radio variability, with the radio peaks
appearing $\lesssim$80 minutes after these weaker X-ray flares. Taken at
face value, these results suggest that stronger X-ray flares lead to longer
time lags in the radio. However, we also test the possibility that the
variability at X-ray and radio wavelengths is not temporally correlated.
We cross-correlate data from mismatched X-ray and radio epochs and
obtain comparable correlations to the matched data.
Hence, we find no overall statistical evidence that X-ray flares and radio
variability are correlated, underscoring a need for more simultaneous, long
duration X-ray--radio monitoring of \sgra.


\end{abstract}




\section{Introduction} \label{sec:intro}

Sagittarius A* (Sgr A*) is a presently dormant supermassive black hole at the
dynamical center of our Galaxy, with a mass $M$ of $\sim 4 \times 10^6$ \Msun\
\citep{Schodel02,Ghez08,Gillessen09}. It has a very low accretion rate
($\lesssim$10$^{-7}$ \Msunyr; \citealt{Marrone06,Shcherbakov12,YusefZadeh15}),
and bolometric to Eddington luminosity ratio ($L/L_{Edd}$ $\sim$ 10$^{-9}$;
\citealt{Baganoff03}). This low $L/L_{Edd}$ can be understood in the context
of a radiatively inefficient accretion flow (RIAF), such as an advection-dominated 
accretion flow \citep[ADAF;][]{Narayan98,Yuan03}. At a distance of $\sim$8 kpc
\citep{Genzel10,Boehle16,Gillessen17}, Sgr A* is a prime target for studies of the
physics and the environment of a low-accretion-rate SMBH \citep{Falcke13}.

For nearly two decades, beginning with the first detection of a flare in the
X-ray \citep{Baganoff01}, \sgra\ has been monitored for episodes of increased
flux.
Variability, which manifests as distinctive flares at high energies, has now
been observed from X-ray
to radio wavelengths.
X-ray flaring detectable by \Chandra\ occurs on average at a rate of 1.0$-$1.3
flares per day \citep{Neilsen13}, although higher rates of X-ray flaring have
been observed \citep{Porquet08,Neilsen13,Ponti15b,Mossoux17}.

With variability occurring at all wavelengths where Sgr A* is detected, it is
instructive to monitor the SMBH simultaneously at multiple wavelengths in an
attempt to detect associated flares in different wavelength regimes and to use
these results to constrain flaring models. These efforts have uncovered
near-infrared (NIR) counterparts for all X-ray flares with simultaneous
observations, although not all NIR flares seem to have X-ray counterparts
\citep[][and references therein]{Morris12}. During these
observations, the corresponding X-ray and NIR flare light curves typically
have similar shapes, and the peaks have measured delays of $<$3 min
(\citealt{Eckart06,YusefZadeh06b,DoddsEden09}; though, see also
\citealt{YusefZadeh12} and Fazio et al. in prep). Their similar characteristics
may point to a common emission mechanism \citep{Witzel12,Neilsen15,Ponti17}.

Light curves at longer wavelengths, i.e., the sub-mm to radio, show different
behaviors, with flares of longer duration delayed by up to a few hours from the
X-ray/NIR flares they are presumed to be associated with. However, existing
observations of simultaneous X-ray and sub-mm/radio flaring are very sparse.
There are three reports in the literature of contemporaneous X-ray and radio
variability.
However, in one case, on 2006 July 17, it is unclear whether the peak in the
radio has been observed \citep{YusefZadeh08}, and, in another, it is not
clear that the X-ray/NIR flare and the radio variability are connected
\citep{Mossoux16}.
\citet{YusefZadeh09} report a simultaneous X-ray/IR flare on 2007 April 4 with
a likely associated radio flare that is
delayed by $\sim$5 hours, but note that the radio observation begins
several hours after the X-ray flare occurs.

In addition to these studies, \citet{Rauch16} detect an NIR flare followed by 
a radio flare 4.5 hours later. There have also been simultaneous observations 
in different sub-mm and radio bands, which point toward longer time lags
with increasing wavelength \citep{YusefZadeh06b,YusefZadeh09,Brinkerink15}.

While the cause of \sgra's observed variability remains an open question,
studies have invoked different models to either simulate flares or provide a
theoretical framework that can account for elements of the observed flaring
behavior. Soon after the first detection
of an X-ray flare from \sgra, \citet{Markoff01} used a jet model to explain
the flaring behavior. Jet models continue to increase in sophistication
\citep{Moscibrodzka13,Moscibrodzka14}, and an adiabatically expanding jet
could explain the observed time lags between `short' (X-ray and infrared)
and `long' (sub-mm and radio) wavelength flares
\citep[e.g.,][]{Falcke09,Rauch16}.
Many models invoke magnetic reconnection as the flare catalyst, followed by
synchrotron radiation and adiabatic expansion \citep{DoddsEden10,Li16}, and,
e.g., \citet{Chan15} also includes gravitational lensing near the event
horizon.
The idea of adiabatic expansion
following a magnetic reconnection event has been expanded upon by
\citet{YusefZadeh06b}, who describe a scenario of an expanding plasma blob,
which can explain the observed time lags, as do the aforementioned jet
models. \citet{Dexter13} present an alternative model where the accretion
disk of \sgra\ is tilted,
and they predict that the NIR and millimeter emission is actually
uncorrelated.

In this work we investigate connections between X-ray and radio variability
with contemporaneous \Chandra\ and {\it Karl G. Jansky Very Large Array}
(JVLA) coverage of Sgr A*, which covered 11 dates in 2013 and 2014. Nine of
these observations yield useful data. We discover a simultaneous strong X-ray
flare and radio rise in one observation, a tentative X-ray flare detection
with clear radio variations in another observation, and several X-ray flares
with tentative radio flux variations. We measure X-ray--radio lags in these
observations and evaluate their statistical significance.
We describe our X-ray and radio observations and reduction in
\S\ref{sec:data}. In \S\ref{sec:detect}, we detail our detections of 
potentially associated X-ray and radio variability, and in
\S\ref{sec:crosscorr} and \S\ref{sec:radcorr}, we investigate the
cross-correlation between the different wavelength regimes. In
\S\ref{sec:discuss}, we perform a ``null hypothesis'' test to assess the
connection between the observed X-ray flares and radio variability. We then
compare our results to previous observations of associated X-ray and radio
variability and discuss how these results fit with theoretical scenarios for
the flaring in \sgra. We conclude briefly in \S\ref{sec:conclusions}.

\section{X-ray and Radio Observations}
\label{sec:data}

Throughout the 2013 and 2014 Galactic Center observing seasons (approximately
March -- October), a number of programs were launched to monitor the \sgra/G2
encounter \citep[e.g.,][]{Gillessen12,Witzel14,Pfuhl15,Ponti15b}. As a part
of an international space- and ground-based effort, we initiated a joint X-ray
and radio campaign with \Chandra\ and the JVLA, and obtained over 30 hours of
simultaneous multiwavelength coverage (successful coordinated observations are
listed in Table~\ref{tab:obssum}). On 2013 April 25 an ultra-magnetic pulsar
(or magnetar), SGR J1745$-$2900, went
into outburst at an angular distance only 2.4\asec\ from \sgra\
\citep[e.g.,][]{Kennea13,Mori13,Rea13,Eatough13,CotiZelati15}.
This new magnetar was the first to be discovered in the vicinity of \sgra, and
focused

\begin{deluxetable*}{lcccccccccc}
\tabletypesize{\footnotesize}
\tablewidth{0pt}
\tablecaption{{\it Chandra} and JVLA Observation Summary} 
\tablehead{\colhead{} & \multicolumn{3}{c}{----------------- {\it Chandra} -----------------} & \multicolumn{6}{c}{------------------------------------------ JVLA ------------------------------------------} & \colhead{} \\
\colhead{Obs Date} & \colhead{ObsID} & \colhead{Obs. Start(UT)} & \colhead{Obs. End}  & \colhead{Proj. ID} & \colhead{Obs. Start(UT)} & \colhead{Obs. End} & \colhead{Config.} & \colhead{Band} & \colhead{Freq.(GHz)} & \colhead{Comment}}
\startdata
2013~May~25    & 15040 &  11:38 & 18:50   & SE0824 &  05:28 & 12:28 & DnC & Q   & 40-48 &  \\
2013~Jul~27    & 15041 &  01:27 & 15:53   & SE0824 &  01:21 & 08:20 &  C  & X   &  8-10 & a \\
2013~Aug~11-12 & 15042 &  22:57 & 13:07   & SE0824 &  00:18 & 07:16 &  C  & X   &  8-10 &  \\
2013~Sep~13-14 & 15043 &  00:04 & 14:19   & SE0824 &  22:08 & 05:07 & CnB & X   &  8-10 &  \\
2013~Oct~28-29 & 15045 &  14:31 & 05:01   & SE0824 &  19:11 & 02:40 &  B  & Ka  & 30-38 & a \\
2014~Feb~21    & 16508 &  11:37 & 01:25   & SE0824 &  11:35 & 19:04 &  A  & Q   & 40-48 &  \\
2014~Apr~28    & 16213 &  02:45 & 17:13   & SF0853 &  07:16 & 14:15 &  A  & X   &  8-10 &  \\
2014~May~20    & 16214 &  00:19 & 14:49   & SF0853 &  05:50 & 13:20 &  A  & K,Q & 18-26,40-50 &  \\
2014~Jul~04-05 & 16597 &  20:48 & 02:21   & SF0853 &  02:33 & 09:32 &  D  & Ku  & 12-18 & a \\
\enddata
\tablenotetext{a}{Poor weather conditions leading to poor atmospheric phase
stability occurred during part of the observation.}
\label{tab:obssum}
\end{deluxetable*}

\noindent
additional interest and observational resources on the Galactic Center.


\subsection{Chandra}

The \Chandra\ observations reported here were centered on Sgr A*'s radio
position (RA, Dec $=$ 17:45:40.0409, $-$29:00:28.118; \citealt{Reid04}).
Most were acquired using the ACIS-S3 chip in FAINT mode with a 1/8 subarray.
The small sub-array was adopted to mitigate photon pileup in the nearby
magnetar and in bright flares from \sgra.
Two observations were performed with different instrument configurations, both
tailored to serendipitous transient X-ray binary observations: (1) 
The 2013 May 25 \Chandra\ observation (ObsID 15040) employed ACIS-S1 through
S4 with the high energy transmission grating (HETG) and a 1/2 subarray to
achieve high resolution
X-ray spectra of the magnetar. JVLA data was also collected on this date, but
since the JVLA observations end just as \Chandra\ observations begin
(Appendix A, Fig.~\ref{fig:app_lc}), we do not discuss these data further in
this work. (2) On 2013 Aug 11-12 (ObsID 15042), we employed the ACIS-S3 chip
in FAINT mode with a slightly larger 1/6 subarray to facilitate coverage of
an outbursting X-ray transient, AX J1745.6$-$2901, located $\sim$1 arcmin
from \sgra\ \citep{Ponti15a}. Since there were no significant X-ray flares
during this observation, the light curves again appear in
Fig.~\ref{fig:app_lc}.

We perform \Chandra\ data reduction and analysis with standard CIAO v.4.8
tools\footnote{Information about the Chandra Interactive Analysis of
Observations (CIAO) software is available at http://cxc.harvard.edu/ciao/.}
\citep{Fruscione06} and calibration database v4.7.2. We reprocess
the level 2 events file with the {\tt chandra\_repro} script, to insure the
calibrations are current, update the WCS coordinate system ({\tt wcs\_update})
using X-ray source positions from \citet{Muno09} and SGR J1745$-$2900
\citep{Rea13}, and extract the $2-8$ keV light curve from a circular region
with a radius of 1\spt 25 (2.5 pixels) centered on \sgra. The small extraction
region and energy filter isolate \sgra's flare emission and
minimize contamination from diffuse X-ray background emission
\citep[e.g.,][]{Baganoff01,Nowak12,Neilsen13} and the
nearby magnetar. The X-ray light curves are shown in Figs~\ref{fig:lc}--\ref{fig:oct13}
and in Fig.~\ref{fig:app_lc} with 300 s bins (green lines) and a typical Poisson error bar (green bar).

\subsection{Jansky Very Large Array}

A total of 11 JVLA observations were taken alongside the \Chandra\
observations, as part of project IDs SE0824\footnote{The last
observation in SE0824 was first attempted in October 2013, but was immediately
interrupted due to a government shutdown. These observations were completed in
February 2014.}
and SF0853\footnote{Two of the observations, in March and April 2014, were
mispointed. The incorrect coordinates for these observations differ from the 
actual coordinates of \sgra\ by an amount larger than the beam size.},
nine of which are suitable for our analysis. The observations were taken
during different configurations of the JVLA, spanning the full range from A
to D configuration, and from X-band to Q-band (8 to 48 GHz). Each of the nine
observations is taken in a single band, except the 2014 May 20 observation,
where the antennae were split between the K-band and the Q-band.

All of the data have been calibrated using the standard JVLA reduction
pipeline for continuum data integrated

\begin{deluxetable*}{lccccccccc}
\tabletypesize{\footnotesize}
\tablewidth{0pt}
\tablecaption{X-ray and Radio Flare Characteristics} 
\tablehead{
\colhead{} & \multicolumn{4}{c}{------------------ {X-ray} ------------------} & \multicolumn{4}{c}{--------------------- Radio ---------------------} & \colhead{} \\
\colhead{Obs Date} & \colhead{Flare Start} & \colhead{Flare Stop}  & \colhead{Duration} & & \colhead{Flare Start (UT)} & \colhead{Flare Stop} & \colhead{Duration} & & \colhead{Delay} \\
\colhead{} & \colhead{(UT)} & \colhead{(UT)}  & \colhead{(minutes)} & & \colhead{(UT)} & \colhead{(UT)} & \colhead{(minutes)} & & \colhead{(minutes)}
}
\startdata
2013~Jul~27    & 03:30 & 03:48 & 17 & &  ? & ? & ? &  & $\lesssim$80 \\
2013~Sep~13-14 & 26:02 & 27:34 & 92 & &  $\leq$26:00 & $\geq$28:56 & $\geq$176 &  & $\gtrsim$125  \\
2013~Oct~28-29 & 16:11 & 16:51 & 40 & &  $\leq$20:14 & ? & ? &  & \textit{$\lesssim$450}\tablenotemark{a} \\
               & 19:56 & 20:12 & 16 & &  $\leq$20:14 & ? & ? &  & \textit{$\lesssim$234}\tablenotemark{a} \\
2014~May~20 & \textit{07:50} & \textit{08:04} & \textit{15}\tablenotemark{b} &  &  07:33 & 09:01 & 88 &  & $\sim$30?  \\
\enddata
\tablenotetext{a}{It is unclear which X-ray flare is associated with the
detected radio variability for \Oct.}
\tablenotetext{b}{The X-ray flare for \May\ is detected only at a low
significance, when the Bayesian Blocks routine is run at $p_0 = 0.39$.}
\label{tab:flares}
\end{deluxetable*}

\noindent
in the \casa\ software
package\footnote{https://casa.nrao.edu/} \citep{McMullin07}. The flux
calibrator is 3C286 (J1331$-$3030) for all observations. The bandpass
calibrator is J1733$-$1304 for all observations except \May, where we use
3C286. The phase calibrator is J1744$-$3116, except for 2013 May 25, where
J1733$-$1304 was used for both bandpass and phase calibration. Throughout
each observation, the array alternated between Sgr A* and the phase
calibrator source. After running the data once through the pipeline, we
carefully inspected the original measurement sets (the raw visibilities)
to identify data that needed to be manually flagged. The
amount of extra flagging varies by observation, and some data require no
extra flagging. If extra flagging was required on the calibration data,
we re-ran the data through the pipeline.

After iterating with the reduction pipeline, we run one iteration of
self-calibration on the \sgra\ visibilities, using the length of the scans as
the solution interval (with the exception of the \Julyb\ observation because
there are very few baselines longer than 50 $k\lambda$). This generally
improves the phase calibration of \sgra, but for most of the observations,
has little effect on the resulting light curves.

\begin{figure*}
\gridline{\fig{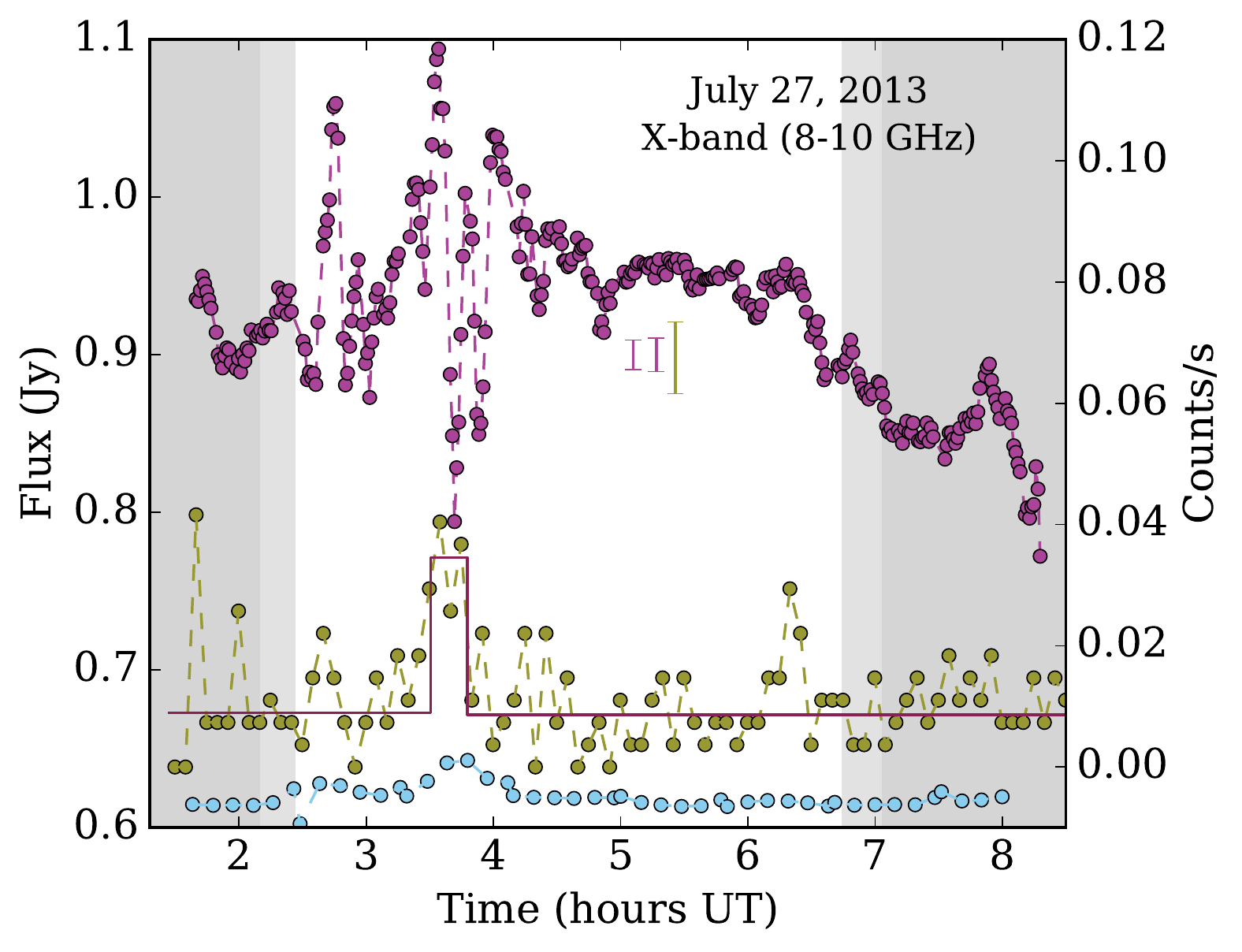}{0.49\textwidth}{}
  \fig{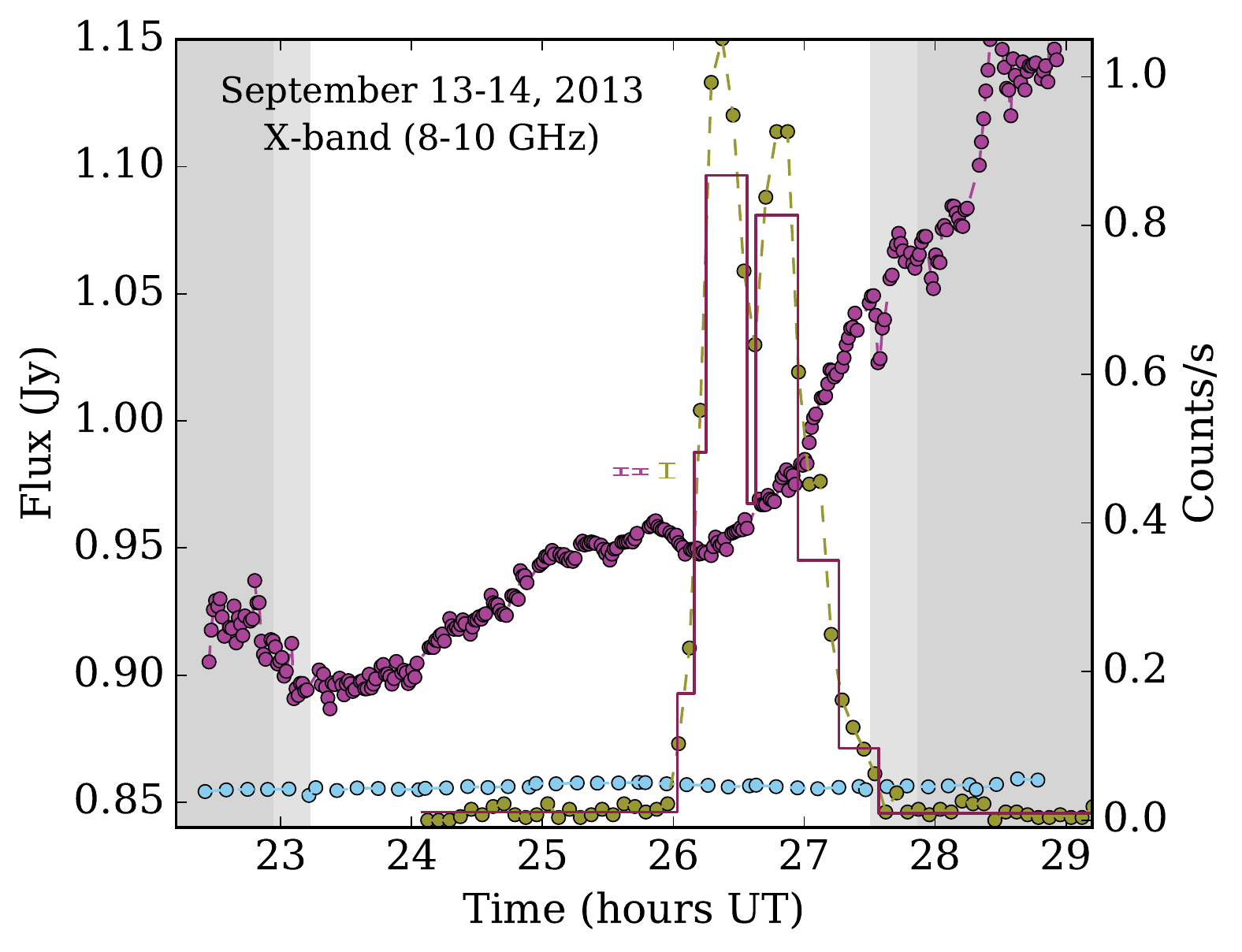}{0.48\textwidth}{}}
\gridline{\fig{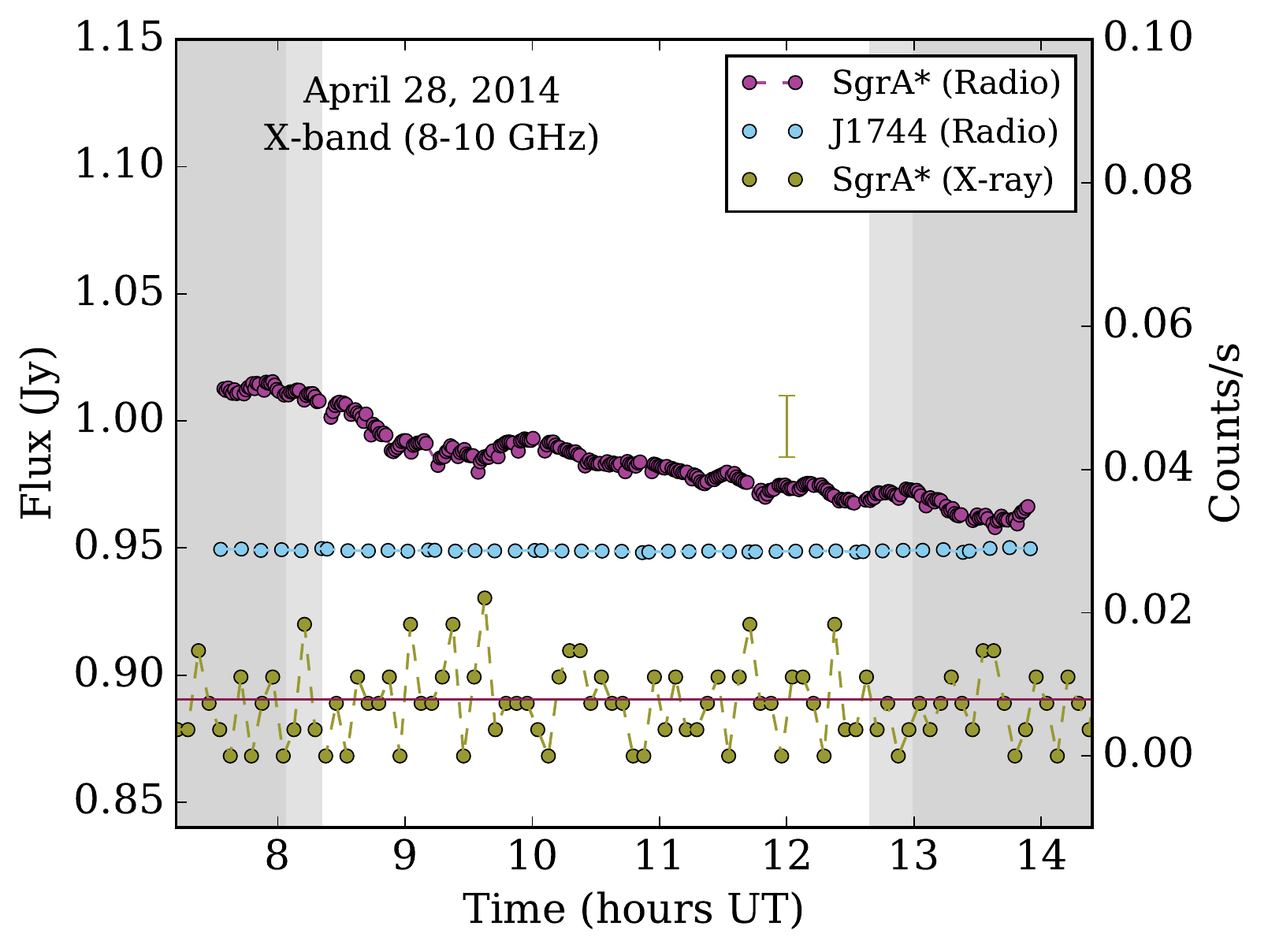}{0.49\textwidth}{}
  \fig{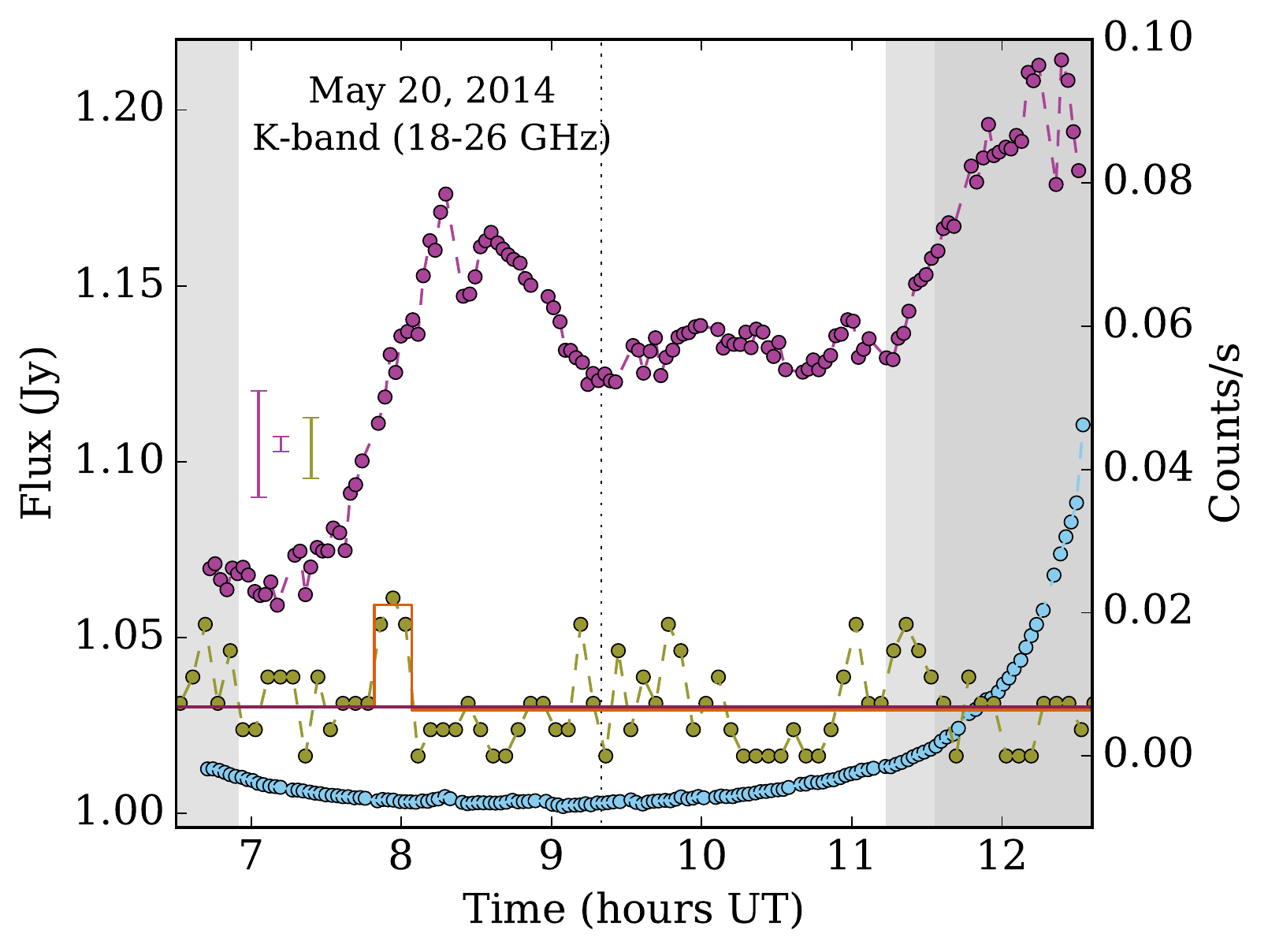}{0.49\textwidth}{}}
\caption{The \sgra\ and phase calibrator radio (purple and blue, respectively)
  and X-ray (green) light curves for three
  observations with significant radio variability (\July, \Sept, and \May).
  The bottom left panel (\Apr) shows a non-flaring radio observation.
  The darker gray shaded regions denote the time intervals where both SgrA*
  and the phase calibrator, J1744$-$3116, were below 18 degrees on the sky,
  and the lightly shaded regions mark times where only J1744$-$3116 is below
  18 degrees.
  The phase calibrator is shifted by a constant factor for display purposes.
  The Bayesian Blocks results are overplotted in red for the X-ray light
  curves; the \May\ light curve includes an additional orange curve
  denoting the lower-confidence Bayesian Blocks results.
  Representative $2\sigma$ error bars are shown for the radio and X-ray light
  curves (for the radio, one error bar is for the entire observation and the
  other for just the higher elevation data in the non-shaded portion of the
  light curve; the error bar for the radio light curve for \Apr\ is too
  small to be visible on the figure).}
\label{fig:lc}
\end{figure*}

\begin{figure*}
\gridline{\fig{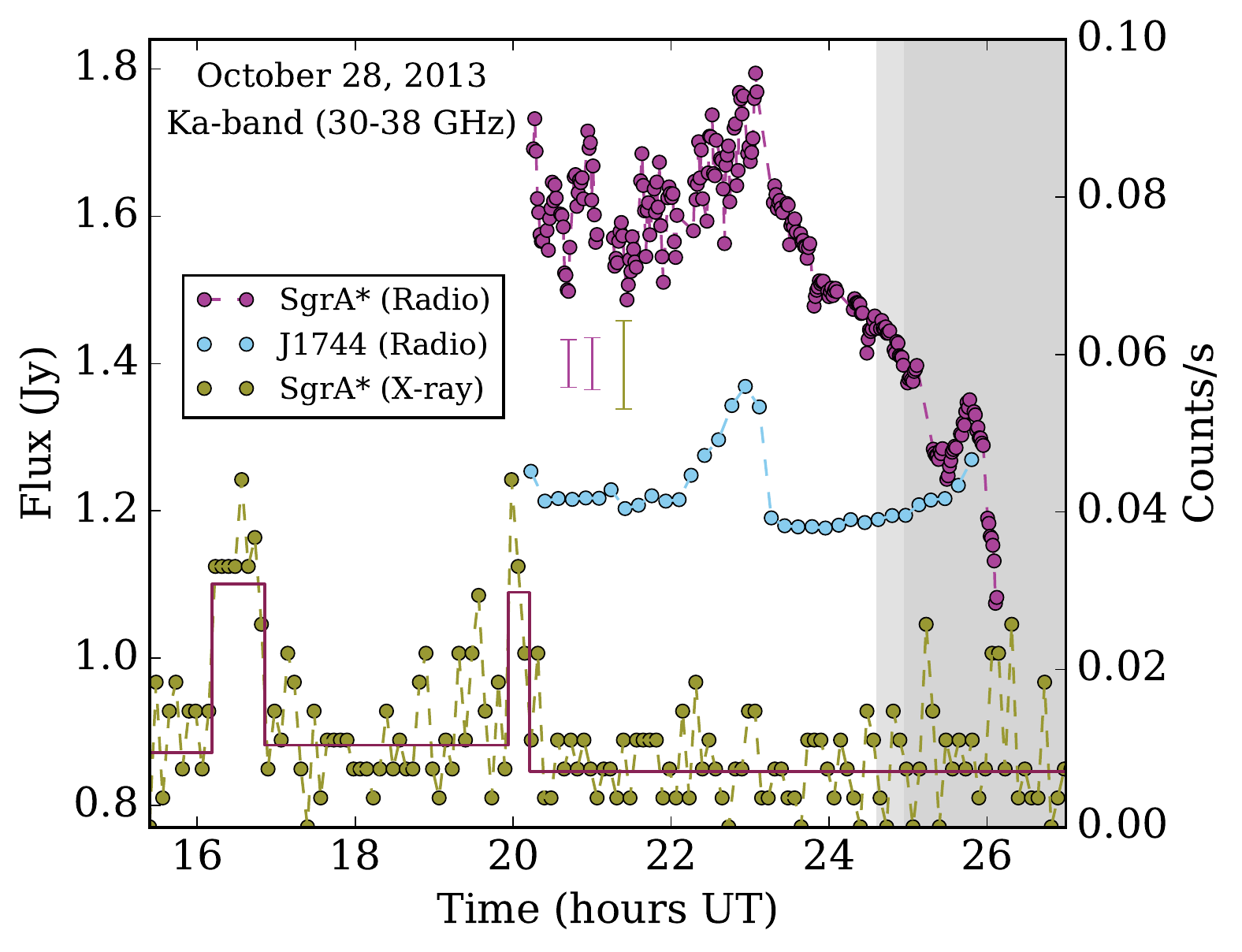}{0.48\textwidth}{}
		  \fig{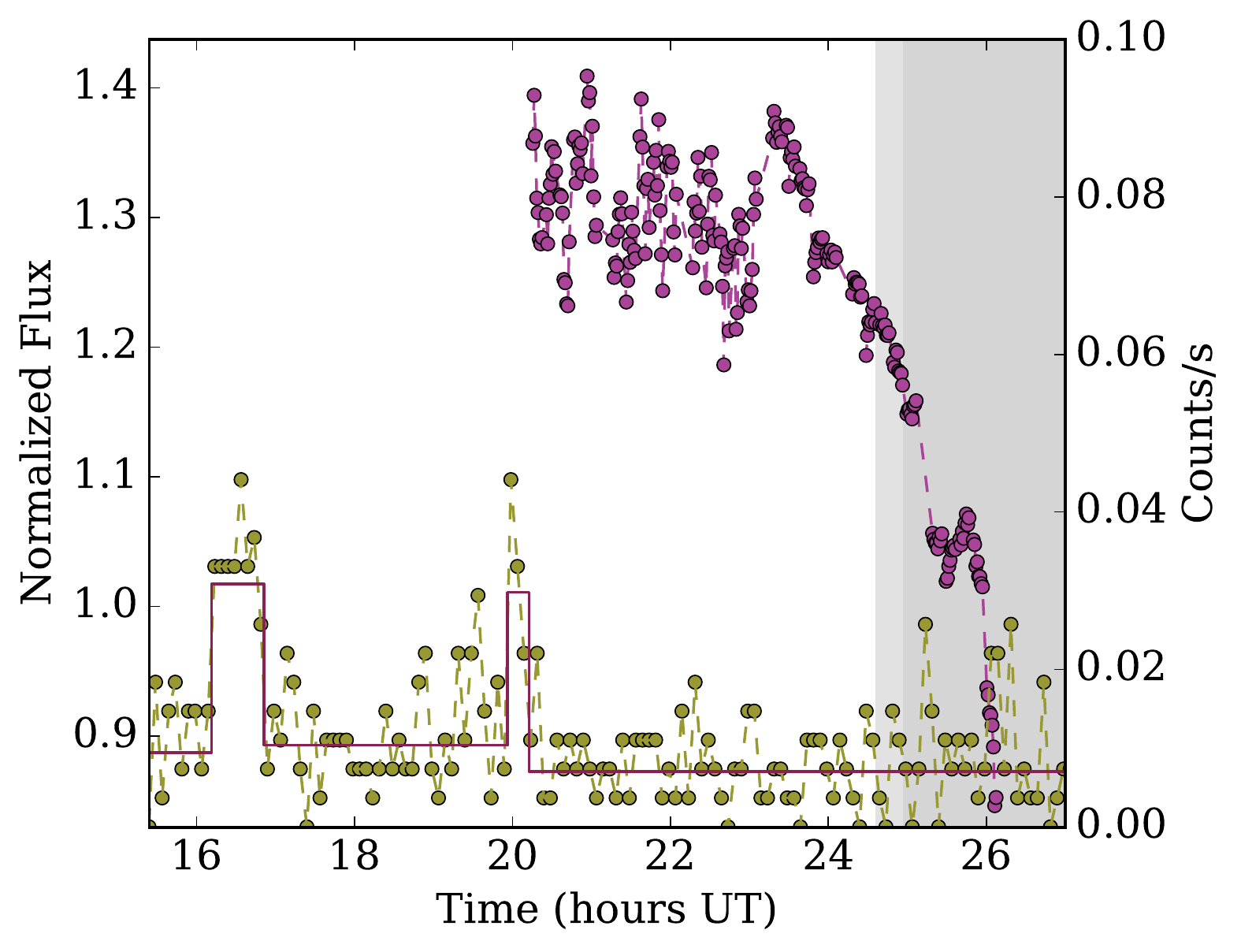}{0.48\textwidth}{}}
\caption{
  Light curves for \Oct, with X-ray in green and radio in purple.
  The blue curve in the left panel is the radio phase calibrator, and the
  Sgr A* radio light curve in the right panel is normalized by the calibrator
  light curve.
  The Bayesian Blocks results are overplotted on the X-ray light curve.
  \label{fig:oct13}}
\end{figure*}

\begin{figure*}
\gridline{\fig{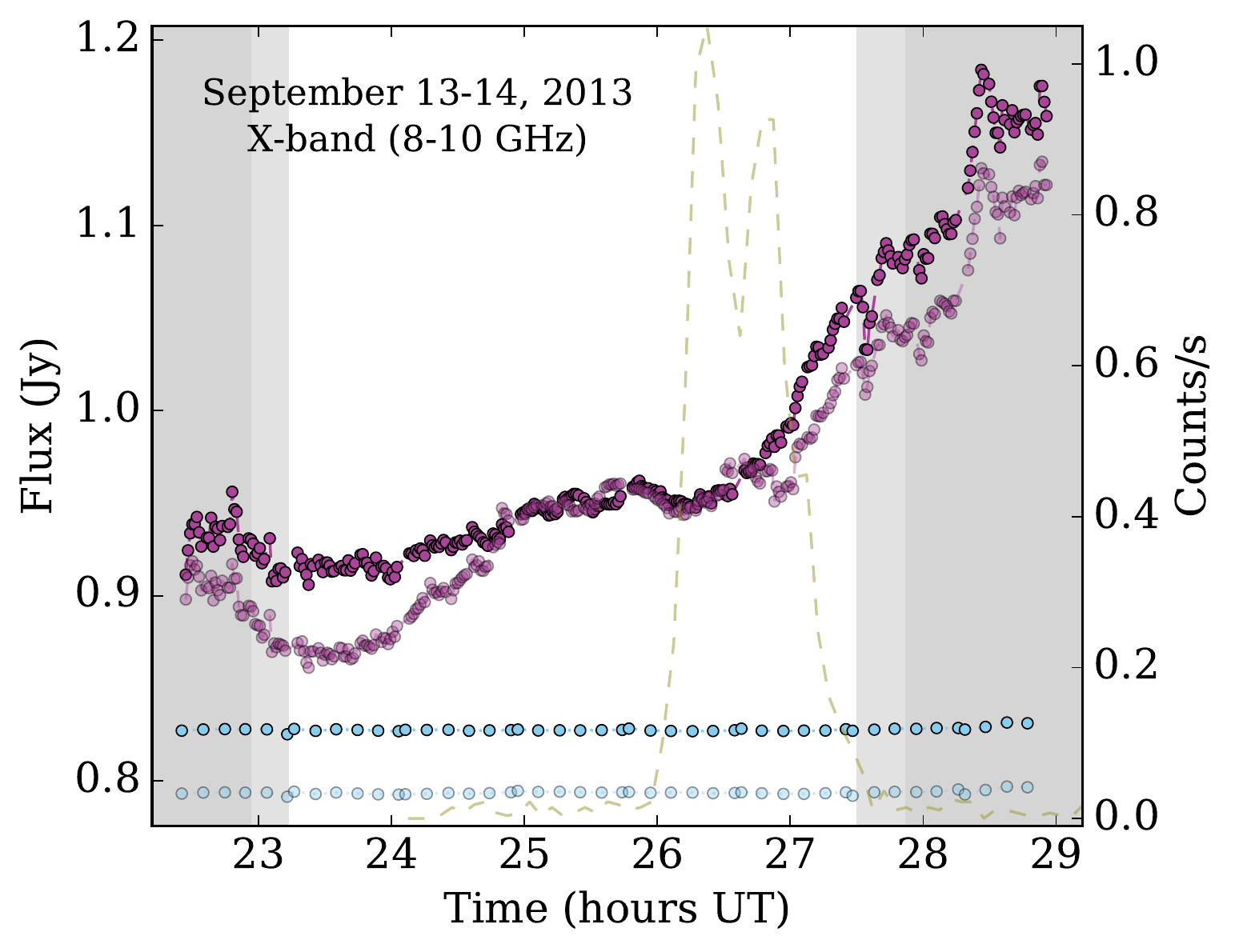}{0.47\textwidth}{}
		  \fig{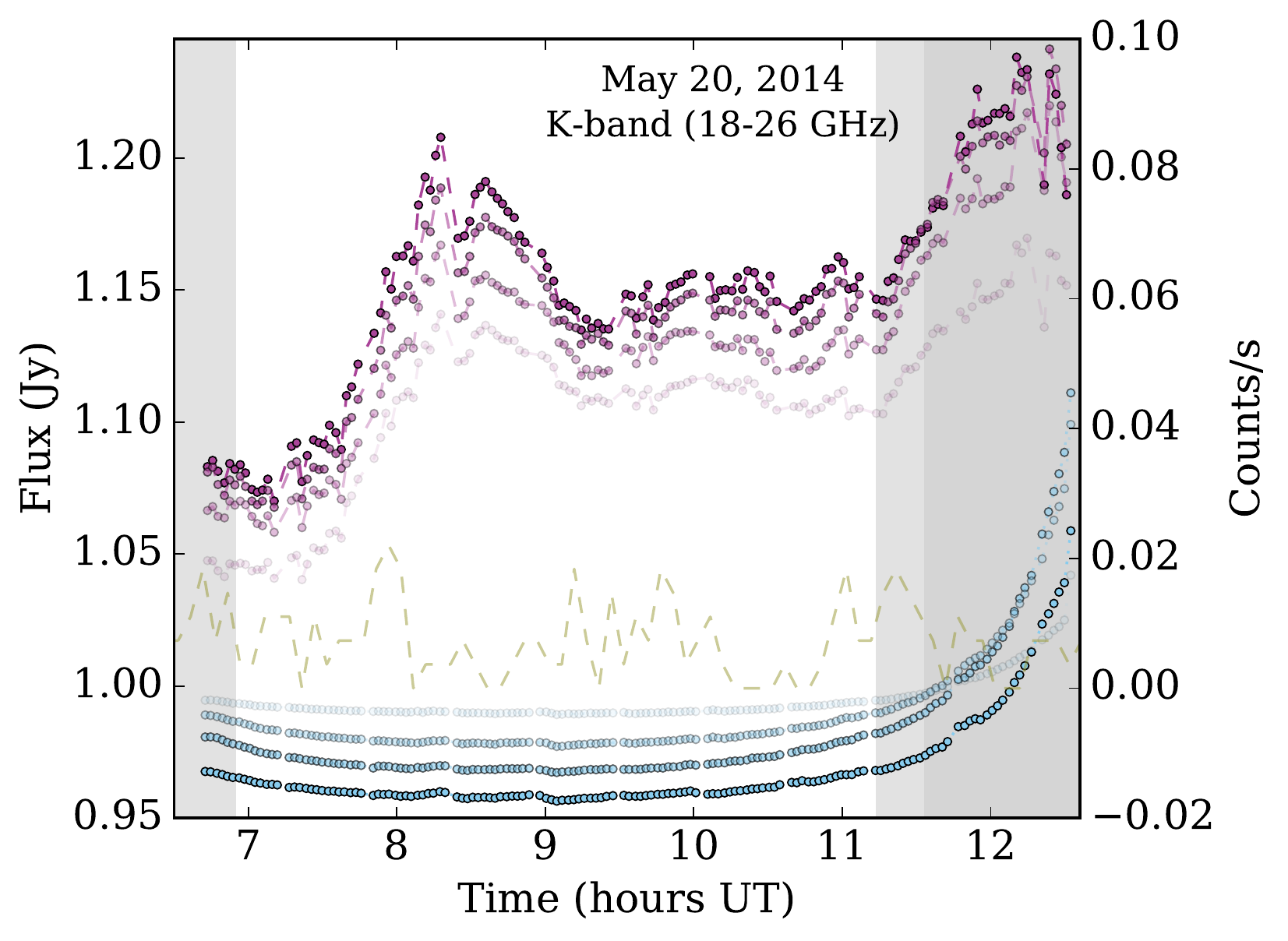}{0.50\textwidth}{}}
\caption{
  Frequency dependence of two radio flares. The darker curves are the highest
  frequencies, and the lighter curves represent progressively lower
  frequencies for each observation.
  Note the dip in the lower frequency bands for \sgra\ around UT 23-24. The
  two light curves meet just before the strong X-ray flare.
  \label{fig:lc_spw}}
\end{figure*}

\begin{figure*}
\gridline{\fig{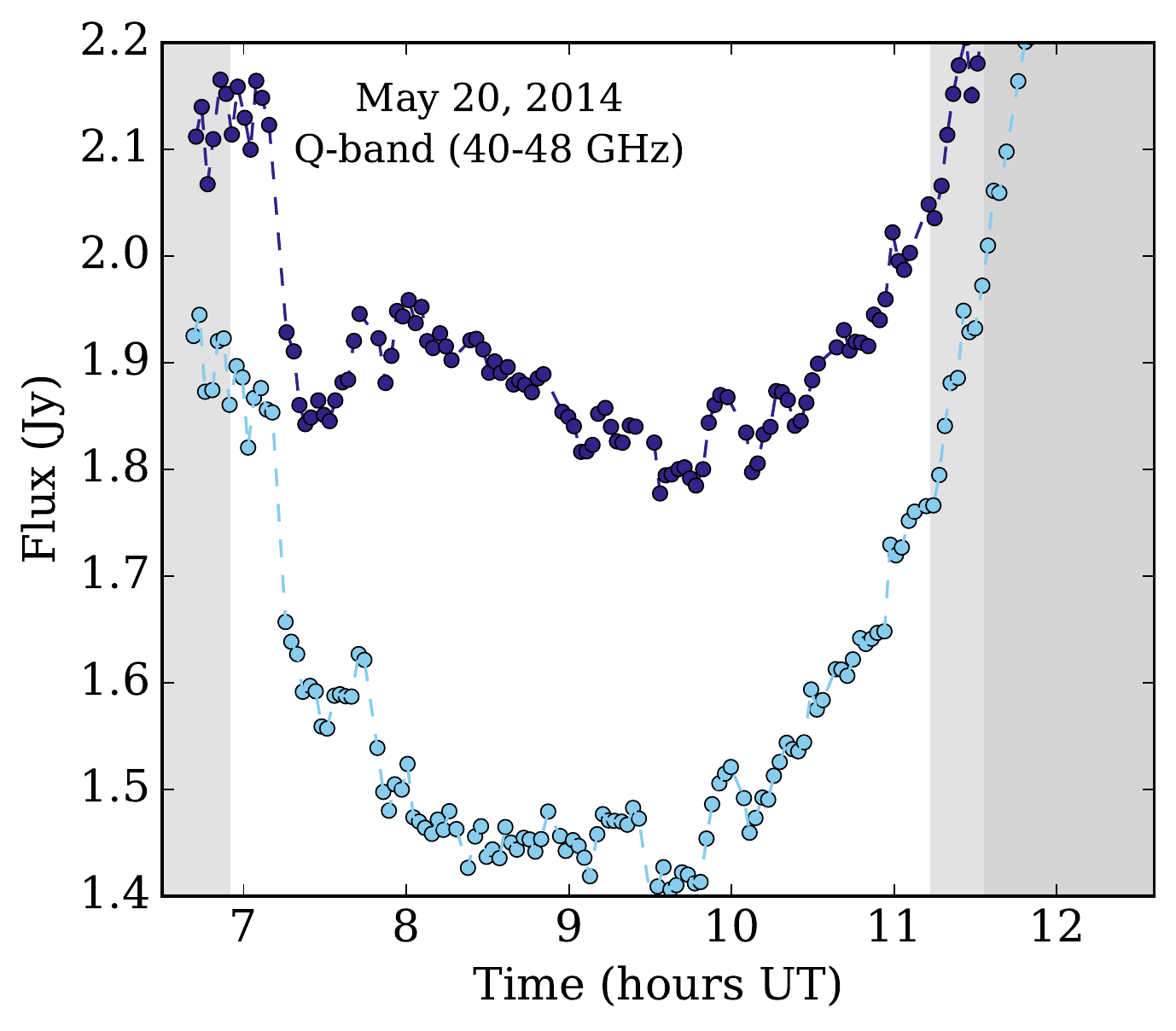}{0.43\textwidth}{}
      \fig{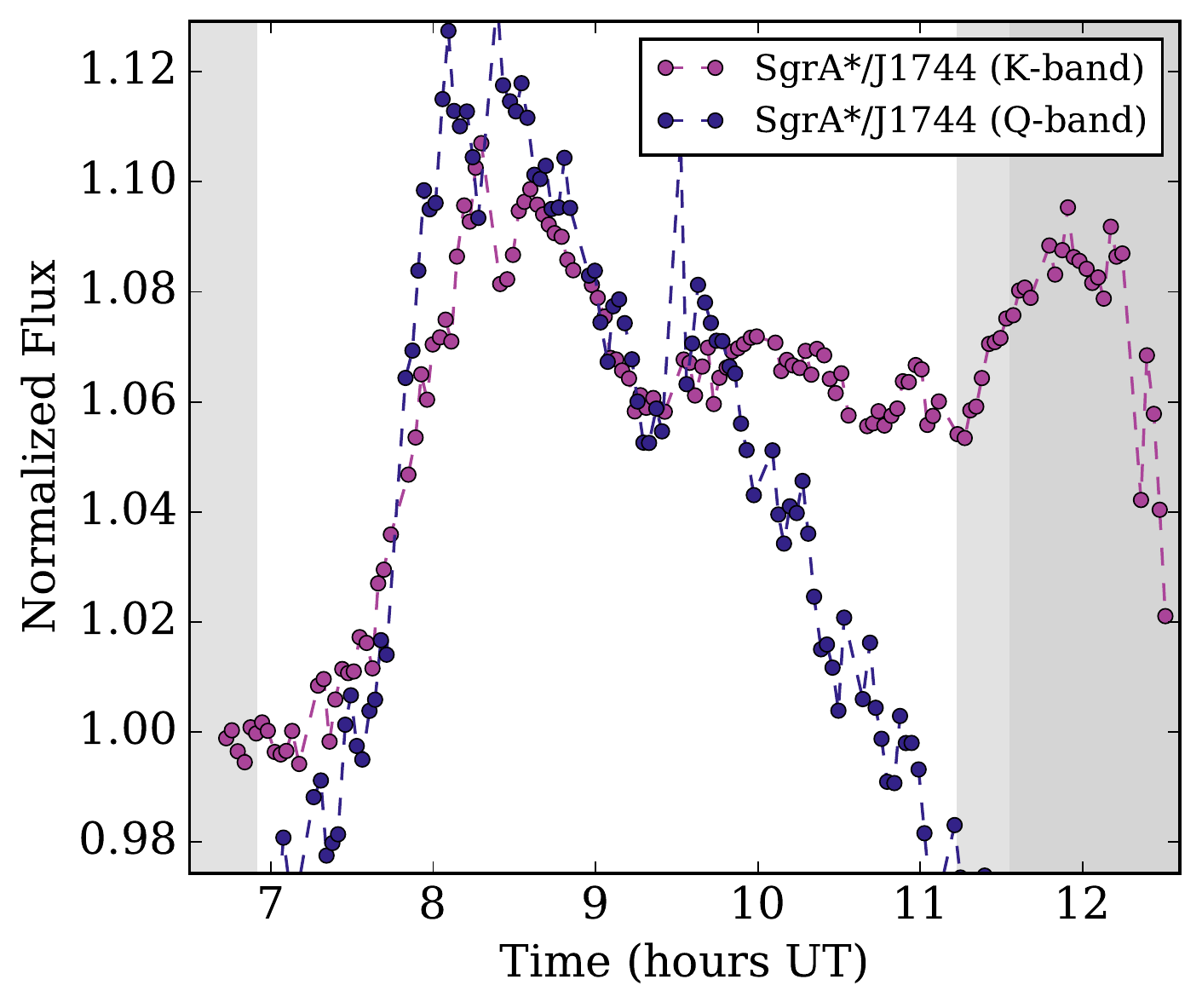}{0.44\textwidth}{}}
\caption{
  The higher-frequency Q-band (43 GHz) observation (dark blue points) for the
  \May\ light curve (left panel) and the light curves for both the K-band
  (23 GHz) and Q-band normalized by the phase calibrator light curve (right
  panel). As in Fig.~\ref{fig:lc}, the phase calibrator is the light blue
  points and the K-band light curve is the purple points.
  \label{fig:may14}}
\end{figure*}

To generate JVLA light curves, we employ the \casa\ command
\textsc{visstat} to calculate the average flux per scan over all antennas and
SPWs for baselines longer than 50 $k\lambda$. This avoids imaging the data
prior to generating the light curves. We independently perform basic imaging
to verify that \sgra\
is at the phase center for each scan and that the 50 $k\lambda$ cutoff does not
include any of the structure on extended scales.
The resulting light curves appear in Figs~\ref{fig:lc}--\ref{fig:may14} and
\ref{fig:app_lc}.

\section{Flare Detection and Characteristics} \label{sec:detect}

\subsection{Bayesian Blocks X-ray Flare Detection}
\label{sec:bblocks}

For X-ray flare detection and characterization, we use the Bayesian Blocks
algorithm
\citep[{\tt bblocks};][]{Scargle98,Scargle13,Ivezic14,Williams17}\footnote{We
adopt the open source Python implementation from Peter Williams, available on
{\it github} at https://github.com/pkgw/pwkit.},
which has been employed effectively in numerous Sgr A* flare studies
\citep{Nowak12,Neilsen13,Ponti15b,Mossoux15,Mossoux16,Mossoux17}.
We run this algorithm with a false positive rate (i.e., 
probability of falsely detecting a change point) of $p_0 = 0.05$. 
This choice for $p_0$ implies that the probability that a change point 
is real is $1-0.05 = 95$\% and the probability that a flare (two change 
points) is real is (1 - $p_0$)$^2$ = 90.25\%.
%
%

We detect only two X-ray flares during the time of overlapping X-ray and radio
coverage, one on \July\ and one on \Sept. We also detect two X-ray flares on
\Oct, but they occur before our radio coverage begins. Similarly, we detect
X-ray flaring activity on \Aug, but it occurs after our radio coverage ends.
The results of the Bayesian Blocks tests are 
overplotted on the X-ray light curves (red lines) in
Figs~\ref{fig:lc}$-$\ref{fig:oct13}~and~\ref{fig:app_lc}.



\subsection{Radio Variability}

\subsubsection{\July}  \label{sec:jullc}

The \July\ observation contains one of the two flares detected in the
X-ray by the Bayesian Blocks routine during the time periods that overlap with
the JVLA observations. The JVLA observations on this date
are unfortunately affected by poor observing conditions, which result in
relatively poorly calibrated phases. Multiple iterations of self-calibration
on the \sgra\ observation do not improve the light curve, but instead indicate
that the structure in the first half of the observation is most likely
spurious. There does appear to be a reliable decline in flux during
the second half of the observation, which may be associated with the detected
X-ray flare.
We plot the \Chandra\ (green) and JVLA (purple) light curves for this
observation in the left panel of Fig. \ref{fig:lc}. The blue light curve is
the radio phase calibrator source.

\subsubsection{\Sept}

The strongest flare in the \Chandra\ data occurs during the \Sept\
observation. The corresponding JVLA data show a clear rise in flux during the
second half of the observation; approximately a 15\% (0.15 Jy) increase. In
comparison, the flux calibrator source, J1744-3116, shows at most a fluctuation
of 0.4\%. We take the standard deviation (0.0014 Jy) of
the calibrator light curve as the typical error for the \sgra\ radio light
curve.

It is difficult to mark exactly where the radio rise begins because we do not
have a long baseline of the quiescent state either before or after the radio
flare.
Before the flare there is some structure in the light curve, including a dip
in the light curve between 23h and 24h. Later, just before the steepest part
of the rise in the radio flux, at around 26h, there is a decline in flux of
approximately 1\% (0.01 Jy). The rise in flux then continues until the end of
the observation.

If we take 26h as the temporal upper limit for the start of the radio flare,
then we have a lower limit on the radio flare duration of $\sim$176
minutes, compared to the full duration of 92 minutes for the X-ray flare.
Whichever time marks the start of the radio flare, it is clear that the radio
rise begins before the X-ray flare begins and the peak occurs after the X-ray
flare ends. A detailed discussion of the X-ray properties of this extremely
bright flare appears in Haggard et al. in prep.

\subsubsection{\Oct}

We detect two X-ray flares in the $\sim$four hours preceding the start of the
\Oct\ radio observation (Fig.~\ref{fig:oct13}). We detect no other X-ray
flares during this \Chandra\ observation. In the radio, the calibration is
poor during the first half of the observation, but there is a clear decline
during the second half of the observation. It is possible that we are
detecting the peak and decline of a counterpart of an X-ray flare.

\subsubsection{\May}  \label{sec:maylc}

While the Bayesian Blocks routine did not detect a significant X-ray flare in
the \May\ observation, we do detect significant structure in the
corresponding JVLA light curve (Figs.~\ref{fig:lc} \& \ref{fig:lc_spw}).
There is a $\sim$9\% (0.096 Jy) increase in flux near the beginning of the
observation, compared to the standard deviation of J1744-3116, which is 0.015
Jy for the entire observation, but just 0.0021 Jy when excluding the very low
elevation data.
For comparison,
we calculate a standard deviation of 0.0048 Jy for the \sgra\ light curve in
the second half of the observation, from 09:20 UT to 11:13 UT (i.e., the
portion of the light curve that appears between the vertical dotted line and
the gray shaded region in Fig.~\ref{fig:lc}), where there appears to be no
significant variability.

We also identify, by eye, a potential weak X-ray flare in the \Chandra\ light
curve around the same time as the rise in the radio.
We therefore run the Bayesian Blocks routine again, relaxing the
$p_0$ parameter. We identify a flare at the time of the radio
rise when $p_0$ is at least 0.39,
indicating that the probability that this
is a real flare is only 37\% (see \S\ref{sec:bblocks}).
We also run the routine with this $p_0$ for the other eight observations and
detect no additional flares (although, the routine does add an additional
block in the \Feb\ light curve at around 13 UT, as shown in
Fig.~\ref{fig:app_lc}).

The rise in radio flux occurs early in the light curve and
remains at 1.07 times the flux value at the start of the observation,
after the flux decreases from its peak value. If the structure in the
X-ray light curve is an actual flare, then it follows a pattern similar to the
\Sept\ flare in that the radio rise begins before the X-ray flare begins and
peaks afterwards.

\subsubsection{Other JVLA Observations} \label{sec:other}

Along with the \Apr\ observation, presented in Fig.~\ref{fig:lc}, the other
four JVLA observations do not show any clear associated X-ray and radio
variability. One of these (\Aug) has an X-ray Bayesian Block detection, but
it occurs after the radio observation ends. We include these light curves in
Appendix A for completeness.

\section{X-ray to Radio Cross-Correlation} \label{sec:crosscorr}

\begin{figure*}
\plottwo{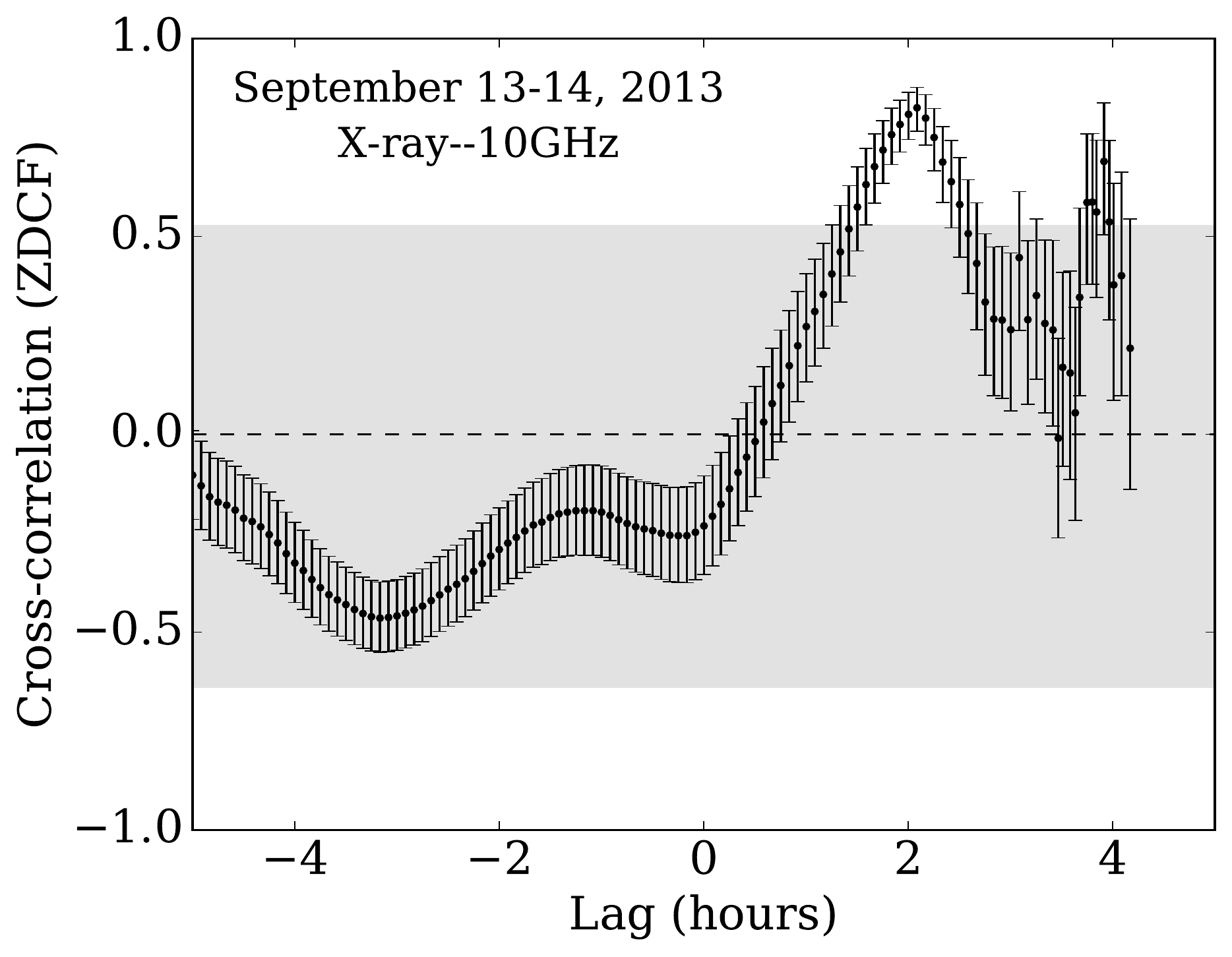}{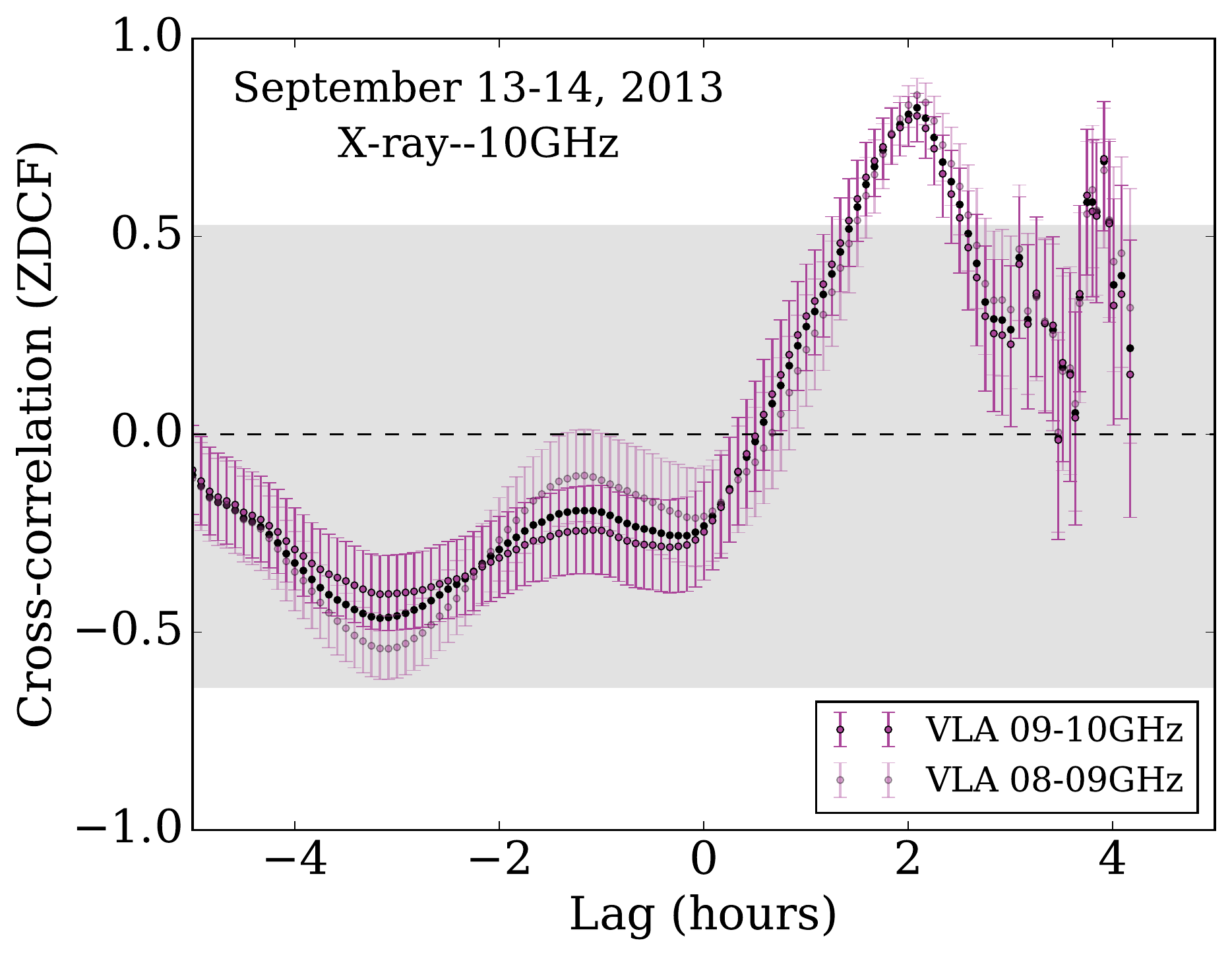}
\plottwo{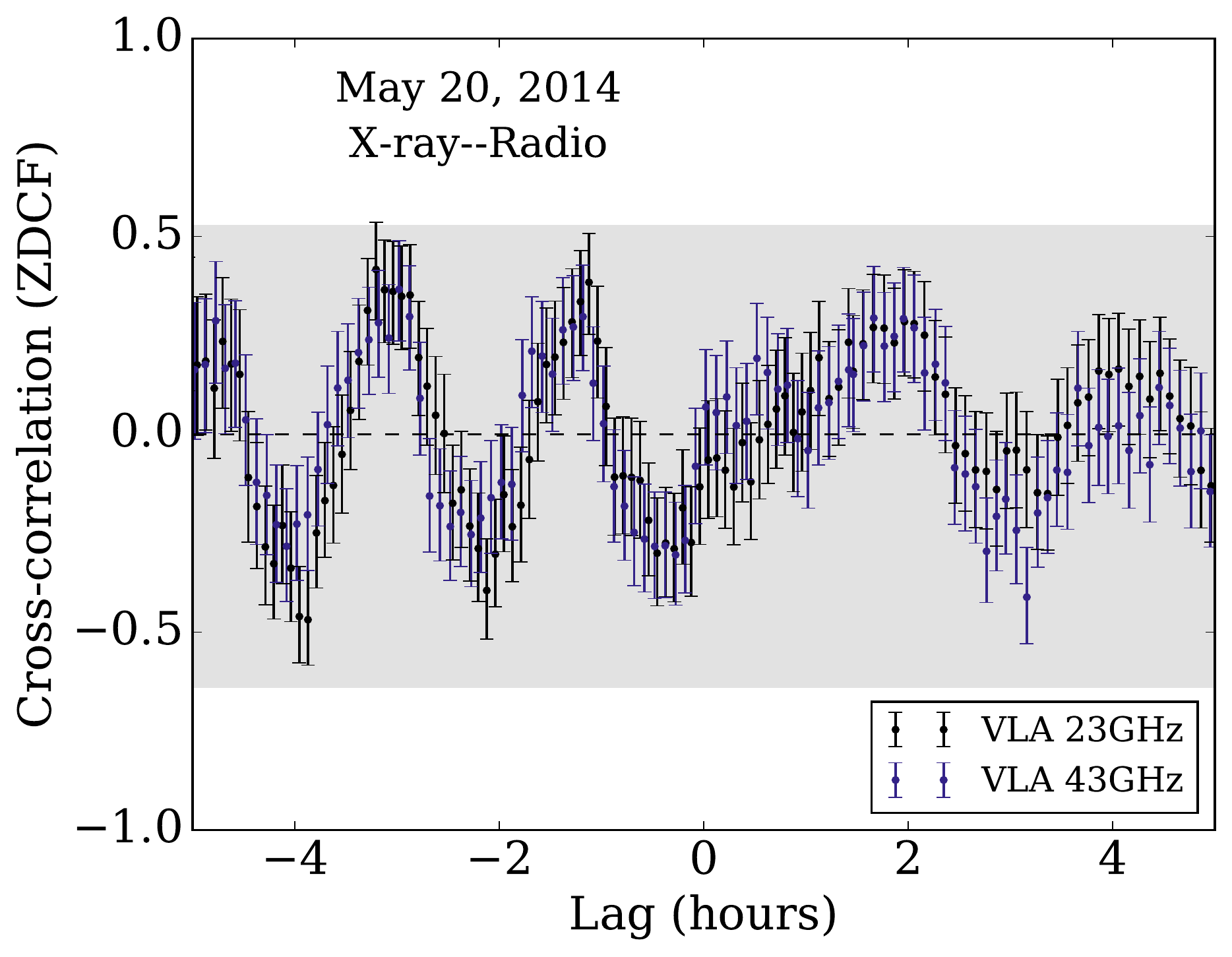}{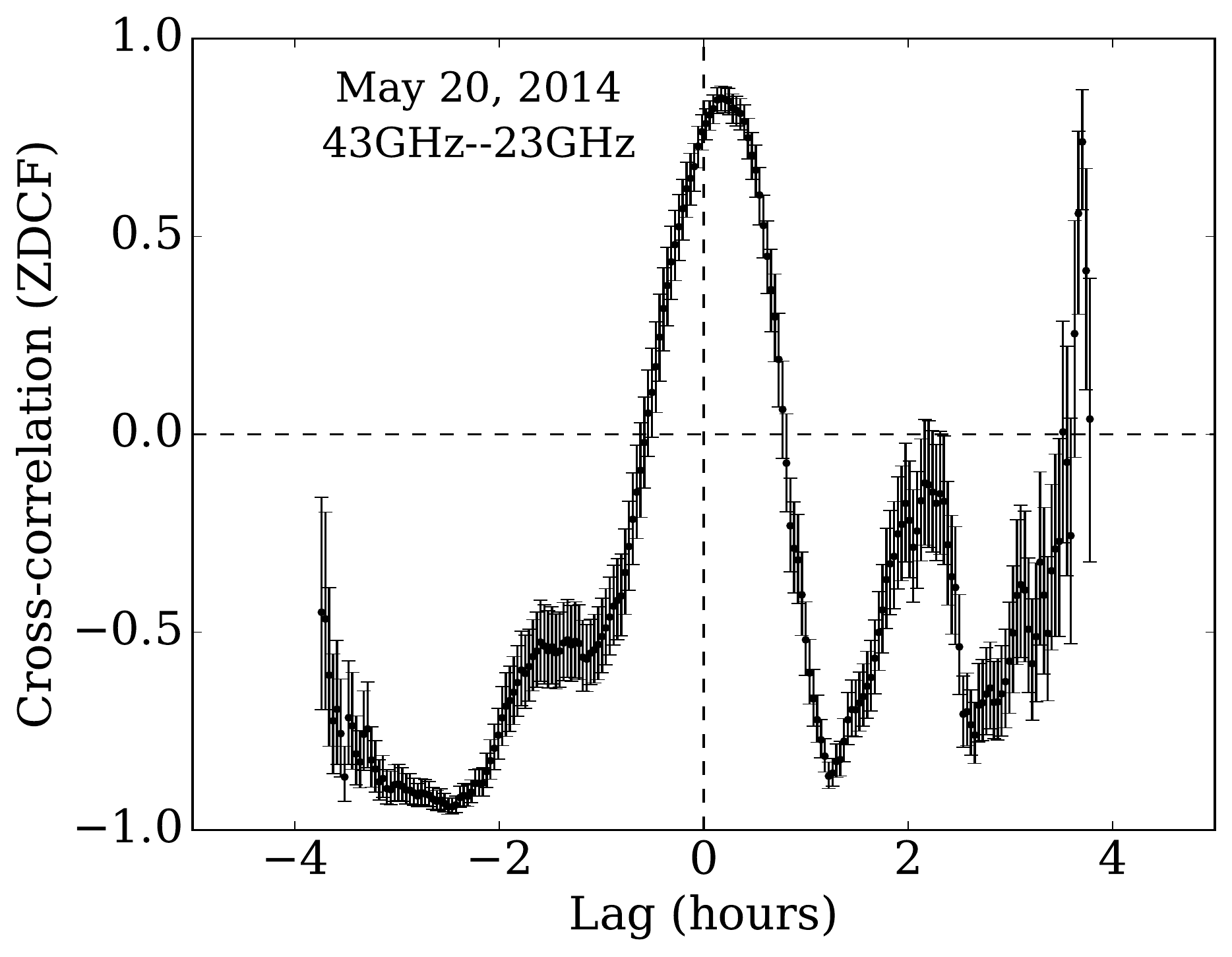}
\caption{The ZDCF for the X-ray and radio light curves for \Sept\ (top
  panels) and for \May\ (bottom panels). The left panels are the
  cross-correlation between the X-ray and the average of all the radio spectral
  windows. The top right panel shows the cross-correlations between the X-ray
  and different radio frequency groupings. The bottom right panel shows the
  cross-correlation between two different JVLA frequency bands for \May. A
  positive lag indicates structure
  appearing in the second data set (radio) after structure appears in the
  first (X-ray). The gray shaded regions denote the 2$\sigma$ error interval
  from the null hypothesis testing (see \S\ref{sec:disc:nht}).
  \label{fig:zdcf}}
\end{figure*}

To quantify the lags between the peaks of potentially associated X-ray and
radio variations, and assess the significance of the lags, we employ the
$z$-transform discrete correlation function \citep[ZDCF;][]{Alexander97}.
Unless otherwise noted, we calculate the X-ray--radio ZDCF using the average
of all the radio spectral window groupings.

\subsection{\July} \label{sec:jul}

Because of the poor calibration for this radio observation, particularly for
the first half where the atmospheric phase stability was poor (see
\S\ref{sec:jullc}), we attempted to fit a smooth polynomial to the light
curve before running the ZDCF. The ZDCF between the X-ray and radio does
not return any significant features (see
Fig.~\ref{fig:mosaic_zdcf}). Despite this, we can estimate an upper limit on
the X-ray-to-radio time lag based on the center of the Bayesian Block flare
detection in the X-ray and the start of the decline in the radio light curve
during the second half of the radio observation, giving a time lag
$\lesssim$80 min.

\subsection{\Sept} \label{sec:sept}

The ZDCF for \Sept\ shows a strong peak, indicating a delay between the peaks
of the X-ray and radio variability, with the X-ray peak leading by $\sim$125
minutes.
The ZDCF is shown in the top left panel of Fig.~\ref{fig:zdcf}.
We also split the spectral windows in half and
calculate the ZDCF using the higher- and lower-frequency spectral window
groupings. Both spectral window groupings give the same strong peak in the
ZDCF (top right panel Fig.~\ref{fig:zdcf}).

However, the radio light curve of this observation continues to increase
until the end of our temporal coverage, indicating that we may not be
detecting the peak of the radio variability. This delay can thus be
considered a lower limit on the time lag between the X-ray and radio, if
they are indeed correlated (\S\ref{sec:disc:nht}).



\subsection{\Oct} \label{sec:oct}

Similar to the \July\ observation, we detect a decline in flux during the
second half of the radio observation for \Oct. In the X-ray, we detect two
flares that precede the radio observation. The X-ray-to-radio ZDCF for the
\Oct\ observation produces several peaks of low significance at
$\sim$2, 4.25, and 7.5 hrs (see Fig.~\ref{fig:mosaic_zdcf}).
We can also use the start of the decline of the radio flux density as an upper
limit on the time of the peak of the potential radio flare, as we do for the
\July\ observation, to obtain an upper limit on any time lag.
If the radio variability is associated with the first X-ray flare,
then the time lag between the X-ray and radio peaks is as high as $\sim$7.5 h,
whereas if the detected radio emission is associated with the second flare,
the lag is less than $\sim$3.9 h.

\subsection{\May}  \label{sec:mayz}

Only very faint X-ray flares, if any, appear in the \May\ observation. Due to
the low significance of a flare detection in the X-ray light curve (see
\S\ref{sec:maylc}), the ZDCF gives several minor peaks (bottom left panel of
Fig.~\ref{fig:zdcf}). In the light curves themselves, there appears to be a
delay of $\sim$30 minutes between the peak flux in the X-ray and in the radio,
with the X-ray leading the radio (bottom right panel of Fig.~\ref{fig:lc}), as
in the \Sept\ flare. The ZDCF instead appears to be identifying weak
correlations between the radio variability at 8h to 9h UT and X-ray
variability between 9h and 10h UT and between 11h and 12h UT.

\section{Radio to Radio Cross-Correlation} \label{sec:radcorr}

For the analyses presented above, we have averaged over all frequencies for
each radio observation. We also investigate the light curves for different
frequency groupings for each observation. Generally, the light curve behavior
is the same in each frequency grouping within a single VLA band, as shown for
the K-band (18$-$26 GHz) observation on 2014 May 20 in
the right panel of Fig.~\ref{fig:lc_spw}.

The 2013 September 14 radio data show some small differences at different
frequencies, in particular, a drop in flux density and recovery in the lower
frequency spectral windows (8$-$9 GHz) at the beginning of the observation.
The peak of the radio variability also appears weaker at these lower
frequencies. We checked
the ZDCF for the X-ray to 8-9 GHz, compared to X-ray to 9-10 GHz, and they
both show a strong peak in the same location (top right panel of
Fig.~\ref{fig:zdcf}).

\May\ is the one observation in our sample with data in two different
frequency bands (K-band, 23 GHz, and Q-band, 43 GHz). This is also one of the
observations with clear structure in the radio light curve, allowing for
comparisons between the light curves in the two different frequency windows.
The lower-frequency, K-band light curve is already presented in
Fig.~\ref{fig:lc}. The higher-frequency, Q-band light curve is shown in the
left panel of Fig.~\ref{fig:may14}, along with the comparison source. To
remove the structure in the light curve that is not intrinsic to SgrA*, we
divide the SgrA* light curve by the comparison source (i.e., the phase
calibrator, J1744$-$3116). For the purpose of
comparing the two radio light curves, we do the same for the K-band light
curve, and the two normalized light curves are presented in the right panel of
Fig.~\ref{fig:may14}.

We calculate the ZDCF for the two radio bands for the \May\ observation and
find a delay between the Q-band (43 GHz) and K-band (23 GHz) of $\sim$10
minutes, with the higher-frequency light curve leading.

\section{Discussion} \label{sec:discuss}

\subsection{Are the X-ray and Radio Correlated?}
\label{sec:disc:nht}

An important caveat to these results
is that while X-ray flares tend to
be distinct events that occur above a fairly smooth, constant background, the
flux at longer wavelengths is almost constantly varying. The radio, in
particular, shows variability at the 8\%, 6\%, and 10\% level at 15, 23, and
43 GHz, respectively, on timescales $<$4 days \citep{Macquart06}. Furthermore,
\citet{Dexter13} predict that \sgra\ has a tilted accretion disk and such a
system could produce variability at millimeter wavelengths that is
uncorrelated with shorter wavelength flaring.

To investigate whether the observed X-ray and radio variability is
actually connected, we run a null hypothesis test to see, for example, if the
radio variability in the \Sept\ light curve is connected to the bright X-ray
flare, or if this could be a chance association of a bright X-ray flare
with typical radio variability.

\begin{figure*}
\centering
\includegraphics[width=175mm]{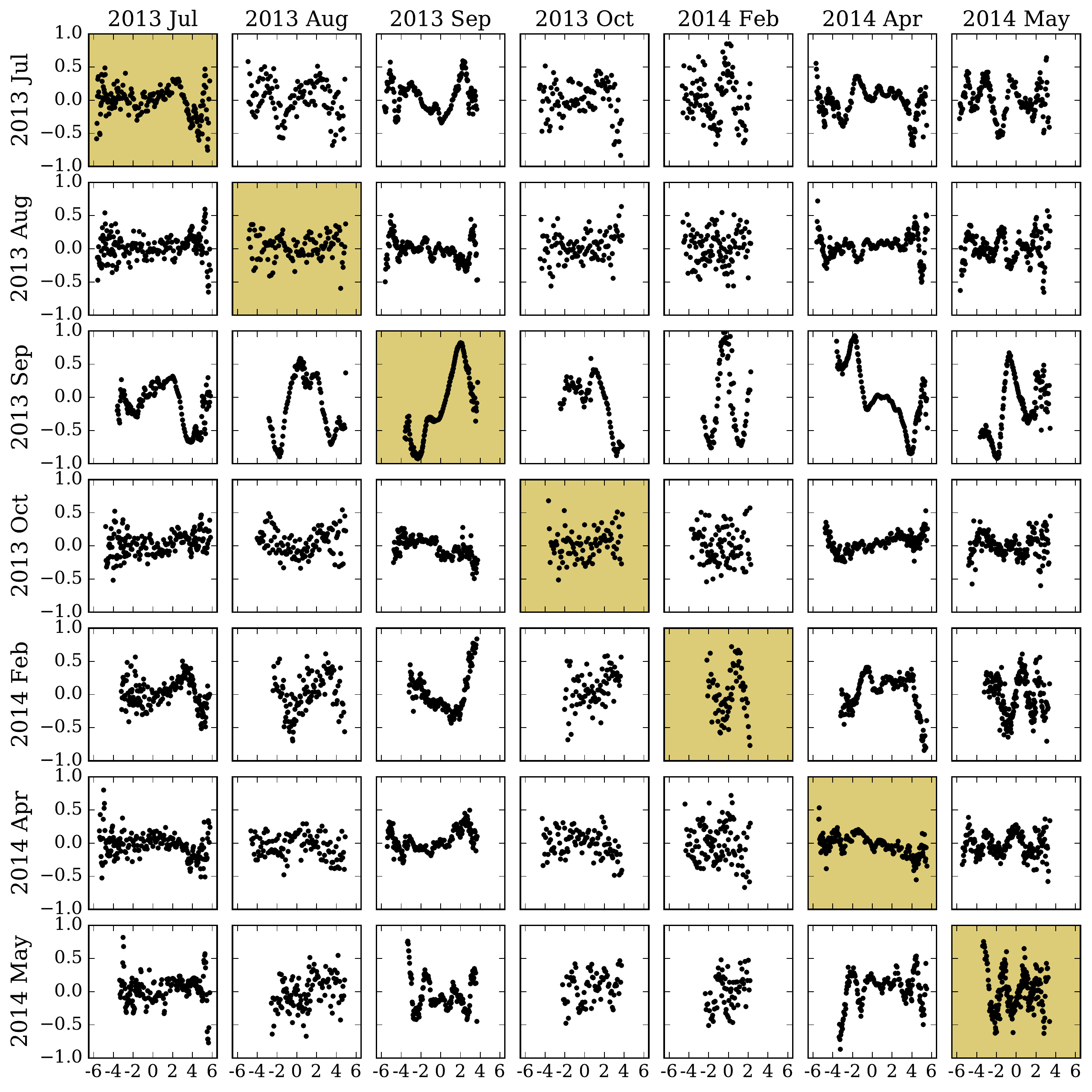}
\caption{The ZDCFs for all pairs of X-ray and radio light curves presented in this paper. The panels along the diagonal, highlighted in yellow, show the correlations for the contemporaneous X-ray and radio observations, discussed in \S\ref{sec:crosscorr}. The off-diagonal panels show the cross-correlation for mismatched X-ray--radio pairs. Although some of the contemporaneous cross-correlations (e.g., \Sept) appear to be significant, many mismatched pairs of observations show comparable correlation peaks.}
\label{fig:mosaic_zdcf}
\end{figure*}

For this test, we utilize all of the X-ray and radio light curves that have
substantial temporal overlap, which reduces our sample to
seven observations. We subtract the UT start time of the radio from both the
radio and X-ray light curves, so that all of the observations start at 0 UT.
We then calculate the ZDCF for each pair of X-ray and radio light curves.
The results are shown in Figure~\ref{fig:mosaic_zdcf} --- there is clear
structure in the ZDCF even when the X-ray and radio light curves are
mismatched. For example, the ZDCFs between the \Sept\ X-ray light curve and
all of the radio light curves show structure similar to the ZDCF between the
matched \Sept\ X-ray and radio light curves. This illustrates that the ZDCF
alone does not prove a physical connection between apparently associated
X-ray and radio variability.

Using the results of this test, we assess the significance of the ZDCF
results presented earlier in Fig.~\ref{fig:zdcf}. We combine the ZDCF values
for all of the mismatched data (i.e., all of the off-diagonal panels in
Fig.~\ref{fig:mosaic_zdcf}), and we calculate the 95.4th percentile
($\sim$2$\sigma$) in both the positive and negative direction. We identify
these error intervals as gray shaded regions in Fig.~\ref{fig:zdcf}.
We note that the errors on the X-ray and radio time series are
non-Gaussian, and thus the errors on the ZDCF may not be statistically
correct. We present them here to provide intuition for the amplitude of the
signal in the ZDCF for uncorrelated \sgra\ X-ray and radio data.
Only the ZDCF for \Sept\ shows a cross-correlation peak outside of this
error interval.

Another test for assessing the possible connection between X-ray and radio
variability is to measure the typical radio variability and compare it to
the radio variability during times of significant X-ray flaring.
The strongest X-ray flare, observed on \Sept, is accompanied by a rise of
$\sim$15\% over $\sim$176 min in the radio at 10 GHz.
This is a larger rise on a much shorter time-scale than the $\sim$8\%
variability found by \citet{Macquart06} for a similar frequency (15 GHz).
The light curve for \Apr\ (Fig.~\ref{fig:lc}) at 10 GHz shows
similar measurement errors,
and has no clear X-ray flares during the X-ray observation, which begins
$\sim$4 hr before and ends $\sim$2 hr after the radio observation.
The \Apr\ radio light curve shows a steady decrease in flux of $\sim$4\%
over $\sim$6 hr, which is a much more gradual change than in the
\Sept\ radio observation.
These results make it less
likely that this is a random radio fluctuation that happens to be
contemporaneous with the X-ray flare. Another radio observation that
may contain a counterpart to a weak X-ray flare, \May, shows
variability at the $\sim$9\% level, which is right at the level of typical
radio fluctuations from \citet{Macquart06}, but again, this is over a
time-scale of $\sim$90 min, which is much less than 4 days.

Future coordinated X-ray and radio campaigns, preferably at a
single radio frequency, would help to establish whether the radio
variability properties differ between times of significant X-ray flaring and
times of X-ray quiescence.

\begin{figure}
\gridline{\fig{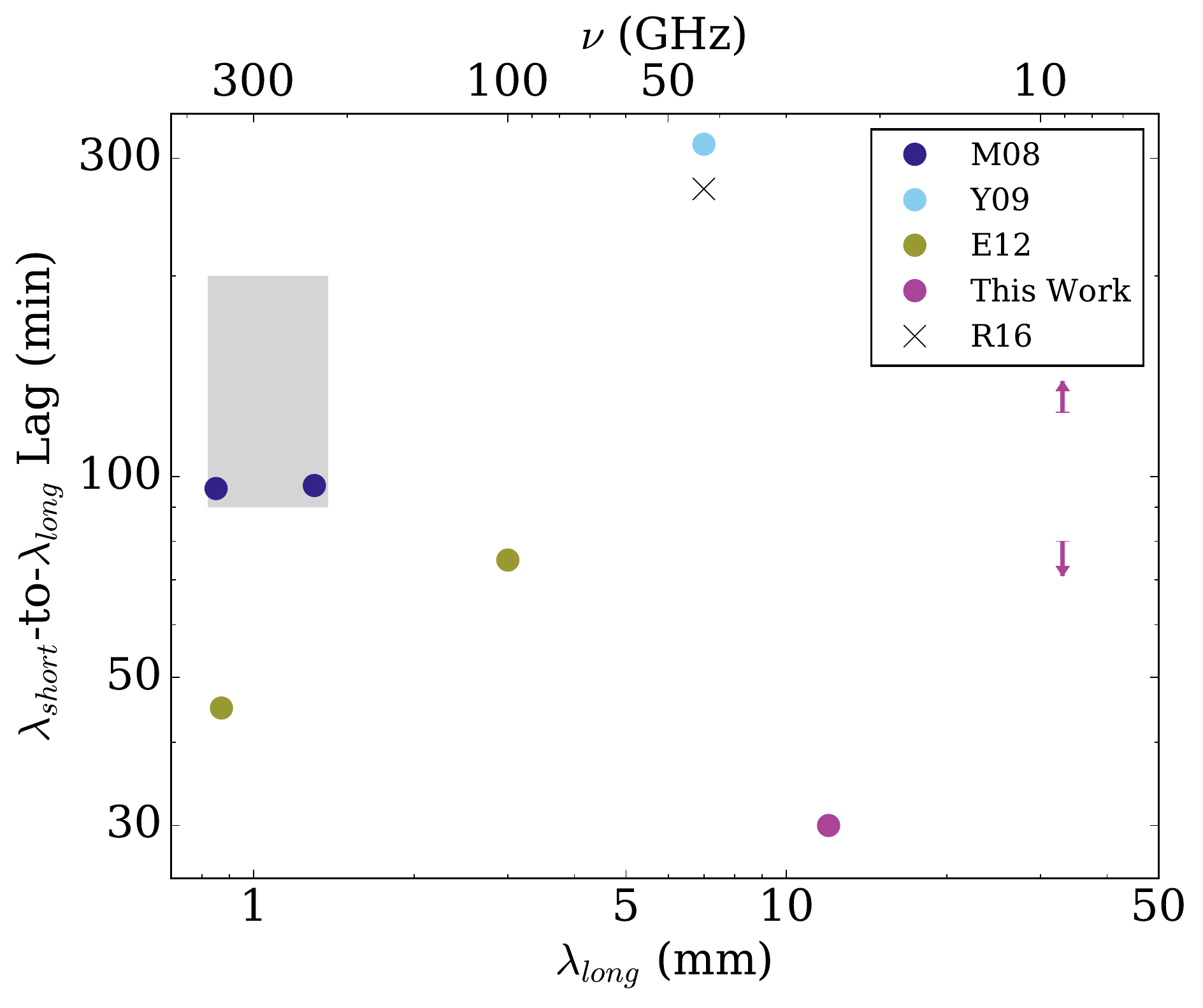}{0.48\textwidth}{}}
\caption{The X-ray or NIR ($\lambda_{short}$) to sub-mm or radio
  ($\lambda_{long}$) time lags for reported flares in the literature and three
  flares reported in this work. The colored points correspond to X-ray
  flares, while the gray box and gray `x' correspond to NIR-only flares.
  \label{fig:lag}
  The references in the legend are \citet[][M08]{Marrone08},
  \citet[][Y09]{YusefZadeh09}, \citet[][E12]{Eckart12}, and
  \citet[][R16]{Rauch16}.}
\end{figure}

\subsection{Comparisons to Previous Work} \label{sec:disc:comp}

Obtaining simultaneous multiwavelength data of \sgra\ flares is challenging,
and only a few previous observations of associated X-ray and submillimeter or
radio flares exist. We collect all of the detections of presumably associated
short- (X-ray and/or NIR) and long- (sub-mm and/or radio) wavelength flares
from the literature, and plot the reported lags, along with our own
detections, in Fig.~\ref{fig:lag}.
We assume that the
short-wavelength flares (X-ray and IR) are simultaneous. However, there
could be delays of a few to tens of minutes between the X-ray and IR, at least
for fainter flares \citep[][Fazio et al. in prep]{YusefZadeh12}.

\citet{Marrone08} and \citet{YusefZadeh08} report submillimeter and radio
variability associated with simultaneous X-ray and IR flares on 2006 July 17.
The peak in the submillimeter occurs $\sim$97 minutes after the X-ray peak
(dark blue points in Fig.~\ref{fig:lag}).
\citet{YusefZadeh08} do not quantify the time lag between the X-ray or sub-mm
and the radio, presumably because the radio peak could have occurred outside
their temporal coverage.

\citet{YusefZadeh09}, on the other hand, finds a delay of $\sim$315 minutes
between an X-ray/NIR flare and a radio flare on 2007 April 4 (light blue point
in Fig.~\ref{fig:lag}), which is significantly longer than the delays we find
here. Even though their radio coverage begins
$\sim$3 hr after the X-ray/NIR flare ends, they argue that the radio flare is
associated with the X-ray/NIR flare because of the high percentage flux
increase compared to their other radio observations on different dates, the
similar morphology between the X-ray and radio light curves, and the absence
of a flare in the sub-mm (240 GHz) during the X-ray/NIR flaring event.

\citet{Eckart12} present X-ray, NIR, sub-mm, and millimeter light curves of a
flaring event on 2009 May 18, with the sub-mm and millimeter flares peaking
about 45 and 75 minutes, respectively, after the simultaneous X-ray/NIR flares
(green points in Fig.~\ref{fig:lag}).

\citet{Mossoux16} detect a rise in the radio at the same time as a flare in
the X-ray and NIR on 2014 March 10. The radio observation ends just as the NIR
flare is reaching its peak, however, and the authors attribute this rise in
the radio to an X-ray/NIR flare that may have occurred before their
observations began.

In addition to these few associated X-ray and sub-mm/radio flares, there are
several reported cases in the literature of associated NIR and sub-mm flares.
The delays between the NIR and sub-mm range from 90 to 200 minutes
\citep{Eckart06,Eckart09,YusefZadeh06a,Marrone08,YusefZadeh08,YusefZadeh09,Eckart08,Trap11}.
There is one report of a lag of just 20 minutes between the NIR and sub-mm
\citep{Marrone08}, but \citet{Meyer08} present additional NIR data showing
that an NIR flare occurred just before the one presented in \citet{Marrone08},
giving a lag of 160 minutes.
We present this general lag time of 90 to 200 minutes between the NIR and
sub-mm as a gray box in Fig.~\ref{fig:lag}.

On 2012 May 17, \citet{Rauch16} detected an NIR flare, followed by a flare at
43GHz, with a lag of 270$\pm$30 minutes. As they lack X-ray information, we
highlight this point with an `x' in Fig.~\ref{fig:lag}.

While the focus of our observing program is to look for correlations between
X-ray and radio flaring activity, we have one observation with simultaneous
observations in two different VLA radio bands (at 43 and 23 GHz).
\citet{YusefZadeh06b} find a delay of 20 to 40 min
and \citet{Brinkerink15} find a delay of 28 $\pm$ 9 min between 43 and 22 GHz.
We find a
shorter time lag of $\sim$10 minutes between these two frequencies (see
\S\ref{sec:radcorr}). \citet{Brinkerink15} also finds a time lag of
$\sim$20$-$40 min between submillimeter and different VLA bands. In these
cases, the lower frequencies are delayed with respect to the higher
frequencies, supporting the trends described above.

From Fig.~\ref{fig:lag}, it is clear that there is not one typical lag time
between shorter wavelength flares and sub-mm/radio variability.
It is also clear that the associations between individual X-ray/NIR and
radio flaring events is uncertain due to the lack of simultaneous data
sampling long time-scales. However, as a whole, these results suggest that
significant lags are common.
If all flares had similar time lags, then one might expect a trend for
increasing lag with wavelength, given the detected lags between sub-mm and
radio frequencies \citep[e.g.,][]{Brinkerink15}. Instead, there is
considerable scatter, and while the longest lags have been found in the radio,
they are not at the longest wavelengths.
This may also support the scenario in which the reported correlations are
spurious (\S\ref{sec:disc:nht}).

For the X-ray$-$radio flares, the two flares with the longest delays, the
2007 April 4 flare from \citet{YusefZadeh09} and the \Sept\ flare from
the current work, have peak X-ray count rates that exceed 1 count/s. These two
are among the brightest X-ray flares known
\citep[e.g.,][Haggard et al. in prep]{Nowak12,Ponti15b}. Because our radio
observations end during the flaring event on \Sept, we do not know when the
radio peak occurs, making it difficult to compare our time lag between the
X-ray and radio with that measured by \citet{YusefZadeh09}.
The X-ray count rates for the two flares with lag times less than $\sim$80 min
(bottom right in Fig.~\ref{fig:lag}; $\sim$0.02$-$0.04 count/s) are much lower
than for the flares with longer lags.

\subsection{Comparisons to Flaring Models}

There are many models in the literature that attempt to explain the flaring
activity of \sgra\ at X-ray through radio wavelengths. Much of the focus is
on the X-ray/NIR flares (and on synchrotron mechanisms), but
several models attempt to explain the sub-mm and radio variations as well.

\citet{DoddsEden10} present a model based on episodic magnetic
reconnection near the last stable circular orbit of the super-massive black
hole, followed by dissipation of magnetic energy.  Their model includes
energy loss via synchotron cooling and adiabatic expansion. The predicted
light curves from their model show simultaneous flaring in the X-ray, NIR,
and sub-mm, with a delayed peak in the radio. They argue that instead of
``flares'' occurring in the sub-mm and radio, there is a decrease in the
sub-mm and radio flux during an X-ray/NIR flare, followed by a recovery.
This recovery then appears as a flare, but is actually a return to
quiescence.
This model does not appear consistent with our X-ray--radio light curves.
In particular, while the \Sept\ radio light curve shows a slight dip at
the start of the X-ray flare, it then shows a significant rise in flux, well
above the flux level in the radio prior to the X-ray flare.

\citet{YusefZadeh06b} adopt the plasmon model of \citet{vanderLaan66} to
explain the observed flaring activity. In this model, there is an adiabatically
expanding ``blob'' of synchotron-emitting relativistic electrons that starts
out optically thick at submillimeter and radio frequencies. Expansion of the
blob's surface area, while the blob remains optically thick, causes the initial
rise of the flare. As the magnetic field decreases in strength, the electrons
cool, and the column density of the expanding blob decreases, the blob becomes
optically thin. At high frequencies, where the blob is initially optically
thin, the model predicts simultaneous flaring and decline in emission. At lower
frequencies, the flaring will be delayed, with the delay increasing with
decreasing frequency.
The predicted lags are consistent with the lags presented together in
Fig.~\ref{fig:lag} and with the lags detected between different
sub-mm and mm frequencies \citep[e.g.,][]{YusefZadeh06b,Brinkerink15}.

Alternatively, if \sgra\ contains a jet
\citep[see, e.g.,][]{Falcke93,Markoff01}, the expanding jet could also
produce time lags between different sub-millimeter and radio frequencies
\citep{YusefZadeh06b,Falcke09,Brinkerink15,Rauch16}. The jet model can
explain the IR to radio spectrum of \sgra\
\citep{Falcke00,Moscibrodzka13,Moscibrodzka14}, and given the observed time
lags, the jet should be at least mildly relativistic \citep{Falcke09}.

General relativistic magnetohydrodynamic (GRMHD) simulations are also being
used to explain \sgra's multiwavelength variability and flaring.
As mentioned in \S\ref{sec:disc:nht}, \citet{Dexter13}
predict that \sgra\ has a tilted accretion disk, within which the emission is
dominated by non-axisymmetric standing shocks from eccentric fluid orbits.
From shock heating, multiple electron populations arise, producing the
observed NIR emission from \sgra, which, in this model, is uncorrelated with
longer wavelength emission.
Alternatively, \citet{Chan15} find that strong magnetic filaments, and their
lensed images near the event horizon, can cause flaring in the IR and
radio, with lags of about 60 minutes. However, their models produce no flaring
in the X-ray, and the predicted lags are slightly smaller than the typical
observed IR-to-submm/mm lags. \citet{Ball16} include a population of
non-thermal electrons located in highly magnetized regions
(also previously considered by \citealt{Yuan03})
and find that X-ray
flares are a natural result. They find cases of simultaneous X-ray and IR
flaring, as well as cases of IR flares with no X-ray counterparts, consistent
with observational results.
\citep[e.g.,][]{Morris12}. However, \citet{Ball16} do not comment on the
emission at submm/mm wavelengths. The MHD model of \citet{Li16} invokes
magnetic reconnection, which leads to energetic electrons that then emit
strong synchrotron radiation. Their model includes the ejections of plasma
blobs, which then leads to the \citet{YusefZadeh06b} explanation of plasma
blobs causing the frequency-dependent delays in observed flaring activity.

Despite intensive focus over the past $\sim$10 years, there is clearly more
work to be done both on the modeling and the observational side of this
problem.
The time lags shown in Fig.~\ref{fig:lag}, if real, indicate that not all
flaring events have the same delay times between the X-ray/NIR and the
sub-mm/mm peaks. This may indicate different physical properties of the
plasma conditions for different events.
Perhaps, if bright flares are caused by a magnetic reconnection event
dissipating a large fraction of the magnetic field into an outflow, in the
process accelerating particles to create a bright flare, there is less
magnetic energy available to accelerate the fluid. Thus, the flow will be
slower than if a smaller flare (or no flare) occurred, leading to a longer
lag time for brighter flares than for weaker flares.
In this work, we also detect an increase in the radio flux before the X-ray
flare begins, which peaks after the X-ray emission returns to quiescence. In
contrast, the X-ray/NIR and radio flares presented in \citet{YusefZadeh09}
indicate a delay of the entire radio flare with respect to the X-ray/NIR
flare, with the radio flare beginning long after the X-ray/NIR flare ends.
More individual multiwavelength flare observations are required to build
a consistent picture of the physics behind these events.

\section{Conclusions} \label{sec:conclusions}

We report 9 contemporaneous \Chandra\ X-ray and JVLA radio observations of
\sgra, collected in 2013 and 2014. We detect a significant radio rise peaking
$\gtrsim$176 min after the brightest X-ray flare ever detected from Sgr~A* on 
\Sept. We also detect a decline in radio flux in two observations (\July\ and
\Oct), following weaker X-ray flares. Finally, we report a significant rise
in radio flux  at the same time as a tentative detection of a weak X-ray
flare, with the radio peaking $\sim$30 min after the X-ray. We thus increase
the number of X-ray--radio correlations from one to 5, where at least an
upper or lower limit on the time lag between the two wavelength regimes can
be measured (Table~\ref{tab:flares}; Fig.~\ref{fig:lag}).

Our study shows trends consistent with previous work on sub-millimeter and
radio variability, which suggest a time lag between an X-ray and/or NIR peak
and the sub-millimeter or radio counterpart (with the long wavelength data
lagging the short wavelength rise). We also detect a time lag of
$\sim$10 mins between the peaks of the \May\ radio variability at two
different frequencies (43 and 22 GHz), again with the longer wavelength
trailing the shorter
wavelength. These results are generally consistent with either an expanding
`blob' or a jet, and combined with the other reported X-ray-to-radio lag in
the literature, our results are suggestive of stronger X-ray flares leading
to longer time lags in the radio.

However, the short durations ($\sim$5 hours) for the radio light
curves makes it difficult to model \sgra's overall radio variability and
limits the utility of cross-correlation functions like the ZDCF.
We perform a null hypothesis test to ascertain whether the correlations
and time lags between the X-ray and radio variations are
statistically significant.
Cross-correlating mismatched X-ray and radio epochs give comparable
correlations to the matched data.
Hence, we find no \textit{statistical} evidence that the X-ray flares and
radio variability are correlated, though our results for the \Sept\ event
remain suggestive. Further monitoring of \sgra\ in the X-ray and radio,
and, in particular, characterizing the radio variability using
non-parametric auto-correlation functions or parametrically, will be
necessary to determine whether there is a physical connection between X-ray
flaring and radio variability.

\acknowledgments

The authors wish to thank the \Chandra\ and JVLA scheduling teams,
in particular Scott Wolk, Pat Slane, and Dillon Foight, for 
their support, which was essential to making these multiwavelength 
observations possible.
We greatly appreciate the kind assistance of the employees behind the
NRAO helpdesk, in particular Drew Medlin and Claire Chandler. 
We also benefited from useful conversations with Mark R. Morris and Nicolas B. Cowan. 
DMC and DH acknowledge support from a Natural Sciences and Engineering
Research Council of Canada (NSERC) Discovery Grant and a Fonds de recherche du 
Qu\'{e}bec - Nature et Technologies (FRQNT) Nouveaux Chercheurs Grant.
ND acknowledges support from a Vidi grant awarded by the Netherlands
organization for scientific research (NWO).
COH acknowledges support from an NSERC Discovery Grant and an NSERC Discovery
Accelerator Supplement.
JN acknowledges funding support from NASA through the Hubble Postdoctoral
Fellowship program, grant HST-HF2-51343.001-A.
GP acknowledges the Bundesministerium f\"{u}r Wirtschaft und
Technologie/Deutsches Zentrum f\"{u}r Luft-und Raumfahrt (BMWI/DLR, FKZ 50
OR 1604) and the Max Planck Society.


\vspace{5mm}
\facilities{Chandra, VLA}

\software{Python, \casa, pwkit \citep{Williams17}}


\appendix

\section{Additional Radio Observations}

In this Appendix, we include light curves for the observations where
there was no clear flare in either the X-ray or the radio during the time of
overlap between the two observations (see \S\ref{sec:other}).

\begin{figure*}[bh]
\gridline{\fig{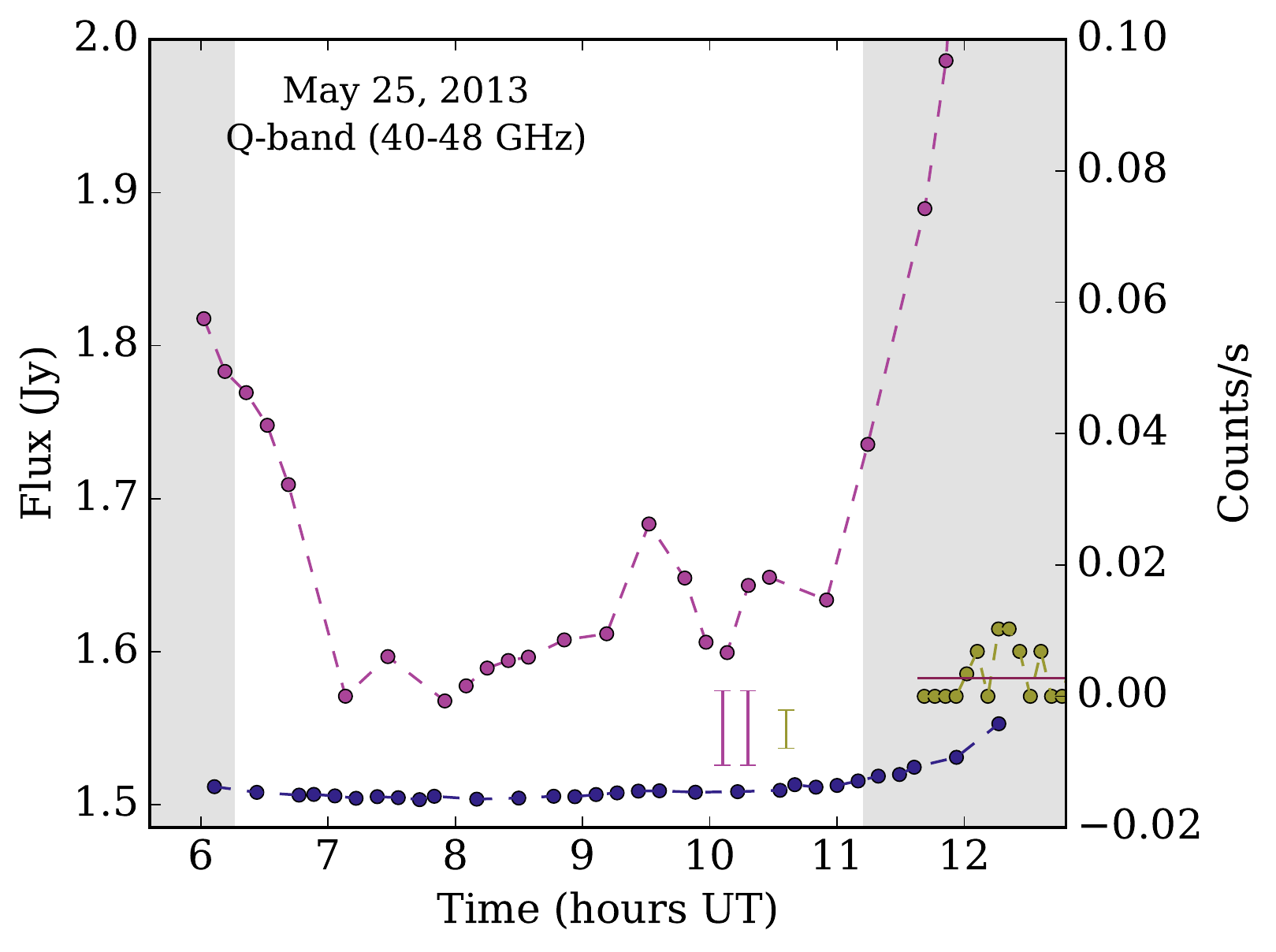}{0.48\textwidth}{}
      \fig{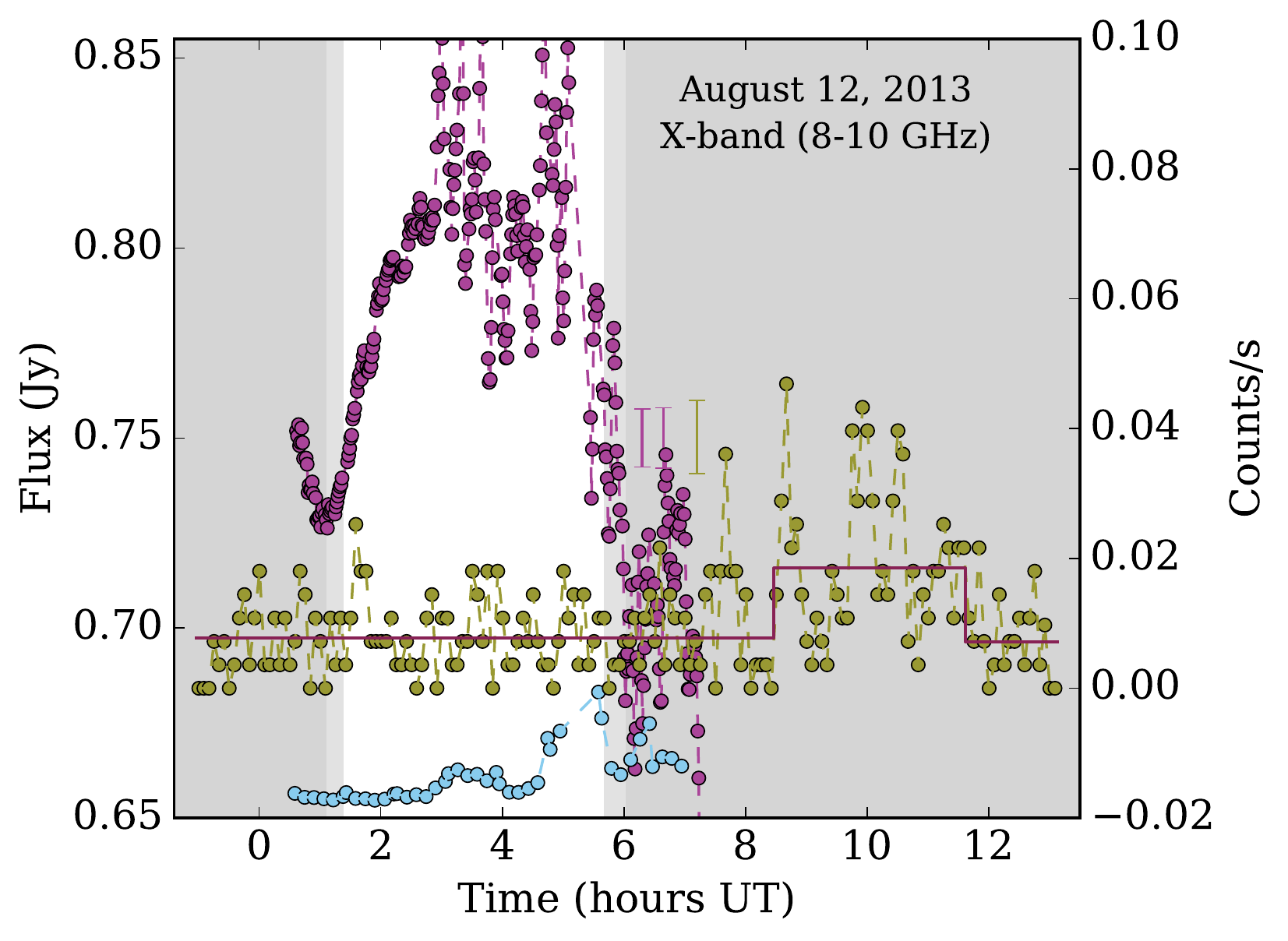}{0.49\textwidth}{}}
\gridline{\fig{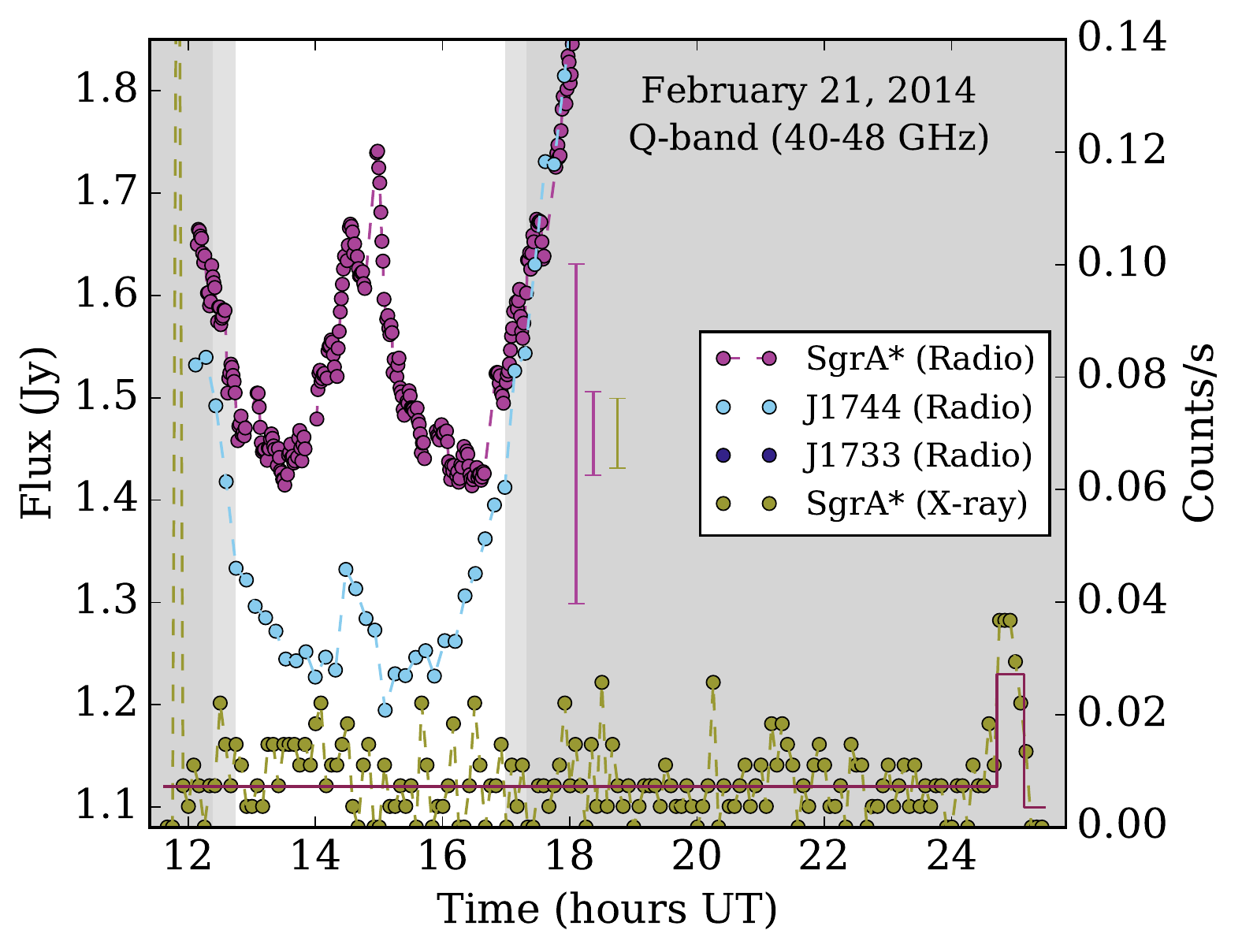}{0.48\textwidth}{}
      \fig{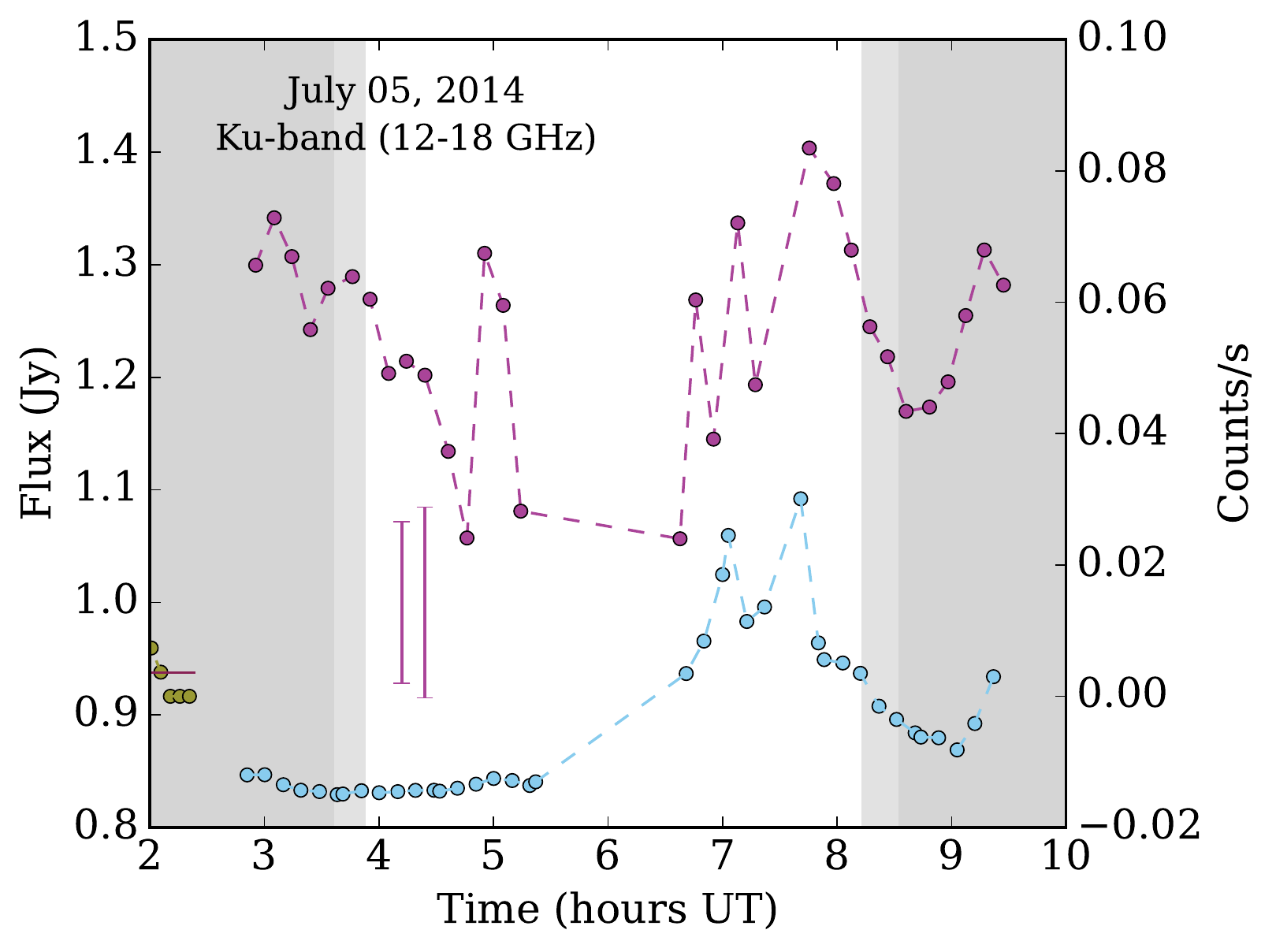}{0.48\textwidth}{}}
\caption{See caption for Fig.~\ref{fig:lc} for details.}
\label{fig:app_lc}
\end{figure*}




\listofchanges

\end{document}
